\documentclass[a4paper,12pt]{article}
\usepackage{amsmath,amsfonts}
\usepackage{graphicx}

\newcommand{\beq}{\begin{equation}}
\newcommand{\eeq}{\end{equation}}
\newcommand{\matrice}{\begin{pmatrix}}
\newcommand{\ematrice}{\end{pmatrix}}
\newcommand{\bea}{\begin{eqnarray}}
\newcommand{\eea}{\end{eqnarray}}

\begin{document}

{\normalsize \hfill SPIN-06/32}\\
\vspace{-1.5cm}
{\normalsize \hfill ITP-UU-06/38}\\
${}$\\

\begin{center}
\vspace{60pt}
{ \Large \bf Quantum Gravity and Matter: \\
${}$\\
Counting Graphs on Causal Dynamical Triangulations}

\vspace{50pt}

{\sl D. Benedetti }
and {\sl R. Loll}\footnote{email:  d.benedetti@phys.uu.nl, r.loll@phys.uu.nl}

\vspace{24pt}

Spinoza Institute and 
Institute for Theoretical Physics,\\
Utrecht University, \\
Leuvenlaan 4, NL-3584 CE Utrecht, The Netherlands.\\


\vspace{48pt}

\end{center}

\begin{center}
{\bf Abstract}
\end{center}
\noindent
An outstanding challenge for models of non-perturbative quantum
gravity is the consistent formulation and quantitative evaluation of
physical phenomena in a regime where geometry and matter are strongly
coupled. After developing
appropriate technical tools, one is interested in measuring and
classifying how the quantum fluctuations of geometry alter the
behaviour of matter, compared with that on a fixed background geometry.

In the simplified context of two dimensions, we show how
a method invented to analyze the critical behaviour of
spin systems on flat lattices can be adapted to the
fluctuating ensemble of curved spacetimes underlying
the Causal Dynamical Triangulations (CDT) approach to quantum
gravity. We develop a systematic counting of embedded graphs 
to evaluate the thermodynamic functions
of the gravity-matter models in a high- and low-temperature expansion.  
For the case of the Ising model, we compute the series expansions
for the magnetic susceptibility 
on CDT lattices and their duals up to orders 6 and 12, 
and analyze them by ratio method, Dlog Pad\'e and differential
approximants. 
Apart from providing evidence for a simplification of the model's analytic
structure due to the dynamical nature of the geometry, the technique 
introduced can shed further light on criteria \`a la Harris and Luck
for the influence of random geometry on the critical properties of
matter systems.  

\vspace{12pt}

\noindent


\newpage

\section{Coupling quantum gravity to matter}

If a quantum theory of gravity is to describe properties of the real world, 
it must tell us if and how matter and spacetime interact at extremely high energies.
Current, incomplete models of quantum gravity usually approach the problem by
first trying to construct a quantum theory of spacetime's geometric degrees of
freedom alone, and then adding matter degrees of freedom in some way. One is
then interested in whether and how the ``pure" gravity theory gets modified and
how the dynamic aspects of quantum geometry may influence the matter
behaviour. A priori, different scenarios at very short distance scales are
thinkable: the behaviour of quantum geometry may be completely dominant, 
geometric and matter degrees of freedom may become indistinguishable, or the
matter-coupled theory may have no resemblance with the pure theory at all.  

In practice, our knowledge of non-perturbative models of quantum gravity, with or 
without matter, and whether they possess physically interesting properties, is at this 
stage limited. However, the development of some of the purely geometric models 
has advanced to a point where one can reasonably consider their more complex
matter-coupled versions, in order to obtain a first quantitative idea of their behaviour.
Such attempts have to overcome a number of serious difficulties. First, there is the
problem of {\it what} to calculate in a background-independent formulation, that is,
to come up with geometric invariants which relate to physical observables. Second,
there is the problem of {\it how} to calculate, once such quantities have been 
identified, that is, the need to establish a viable calculational scheme. In case one uses
a numerical approximation, say, to evaluate a non-perturbative path integral over 
geometries by Monte Carlo methods, one has to deal with the computational
limitations of the hardware. In case one uses some expansion scheme, it must
have non-standard features in order to be inequivalent to the usual 
(non-renormalizable) gravitational perturbation series.

Many difficulties stem from the fact that in the case of quantum gravity, conventional
calculational methods have to be adapted to a situation where there is no fixed
background spacetime, and instead ``geometry" is among the dynamical degrees of
freedom. In this paper we will look in detail at an example of such a generalized method
for a particular class of quantum systems of gravity and matter. We will work in the 
simplified context 
of two spacetime dimensions, where it is fairly clear {\it what} one would like to calculate, 
and where the computational limitations are minimal compared with the full,
four-dimensional theory. More specifically, we will look at spin systems coupled to
dynamical Lorentzian geometries, given in the form of an ensemble of triangulated spacetimes
in the framework of {\it Causal Dynamical Triangulations (CDT)} (see \cite{reviews} for
recent reviews).   

We will show how a time-honoured method of estimating the critical behaviour of a
lattice spin system, namely, by considering finitely many terms of a weak-coupling
(i.e. a high-temperature) expansion of
its thermodynamic functions can be adapted to the case of ``quantum-gravitating" 
lattices. The method consists in classifying local spin configurations (which can be
represented by diagrams consisting of edges or dual edges of a triangulated
spacetime), and determining their weight, i.e. the probability to find such a configuration 
in the ensemble of dynamical triangulations. Diagrammatic techniques have also
been employed recently in another coupled model of geometry and matter in three
dimensions, in the context of a so-called group field theory, an attempt to 
generalize matrix model methods to describe ensembles of geometries in
dimension larger than 2 (see \cite{oriti} for a recent review). 

The model we will be studying in detail is that of an Ising model whose spins
live on either the vertices or triangles of a two-dimensional simplicial lattice
and interact with their nearest neighbours.
Summing over all triangulated lattices in the usual CDT ensemble, each with
an Ising model on it, gives rise to the matter-coupled system whose properties
we are trying to explore. 
An exact solution to this Lorentzian model has not yet been found. This has to do with
the fact that in terms of the randomness of the underlying geometry the model
lies in between that on a fixed regular lattice and that on purely 
Euclidean triangulations,
but is sufficiently different from either to make the exact solution methods known
for these cases inapplicable. 

What underlies the successful application of powerful matrix model
methods to solve the model of {\it Euclidean} dynamical triangulations with Ising
spins \cite{kaz} is the fact that on Euclidean geometries, no
directions are distinguished locally, which matches with the fact that the matrix 
model generates configurations
(in the form of triangulated surfaces or their dual graphs) which are {\it unconstrained}
gluings of its elementary building blocks. 
However, it turns out that for the purposes of quantum gravity
this property of local ``isotropy" is simply not good enough. Ensembles of 
unconstrained gluings of triangular building blocks (so-called four-simplices) in
four dimensions, even if they are required to be manifolds of a fixed topology,
have been found to be dominated by highly degenerate polymeric or tightly
clustered geometries, neither of them making good candidates for a theory of
quantum gravity (see \cite{4drev} for a review). 
This was precisely the reason for introducing the
Lorentzian CDT version of the original (Euclidean) dynamical triangulations,
which does keep track of the non-isotropic light-cone structure and flow of
time on its spacetimes. Remarkably, in this model, evidence for the desired
four-dimensionality of its geometric
ground state on large scales has recently been found \cite{ajl-prl,semi,ajl-rec}. 
In more recent times, the introduction of somewhat different ``causality" constraints 
has been advocated in a Euclidean spin-foam approach to quantum gravity
(see \cite{ort} and references therein).

Returning to our previous argument, it is notoriously
difficult to introduce additional restrictions concerning the local nature of
geometry into a matrix model without destroying its simplicity and, more importantly,
its solubility.
This also seems to apply to the two-dimensional CDT model, whose pure-gravity version
was first solved exactly in \cite{al} by transfer-matrix method, with the Ising-coupled 
model studied subsequently in 
\cite{aal1,aal2}.\footnote{We would expect a similar issue to appear in the group 
field theory models mentioned 
earlier, which beyond triangulated manifolds generate a vast number of geometric 
configurations, many of which one presumably would like to get rid of by
imposing suitable constraints (see also the remarks in \cite{oriti}).} 
In the latter case there is strong evidence from numerical simulations that
the spin model behaves in the presence of gravity (at least as far as
its critical exponents are concerned) just like it does on a fixed, regular lattice.
This is supported by a ``semi-analytical" high-temperature 
expansion, whose initial results were merely quoted in the original work of \cite{aal1}.
In the present paper, we
give a detailed technical exposition of how to perform the relevant graph
counting on causal dynamical triangulations, both for the standard Ising 
model and its dual. The algorithms found have allowed us to compute
the diagrammatic contributions to the Ising susceptibility to
order 6 and order 12 for CDT lattices and their duals, respectively.

The pure counting results were announced in a recent letter \cite{ising1}, where
we put forward the conjecture that the best way to extract information about
the universal properties of a matter system on {\it flat space} is by coupling
it to an ensemble of quantum-fluctuating causal triangulations! 
The strongest evidence for this conjecture so far comes from the fact that an evaluation by
straightforward ratio method of a rather small number of susceptibility coefficients 
already leads to results in surprisingly close agreement with the exactly known 
value for the susceptibility. As we will also reiterate in the present paper, the
evidence for this conjecture is still too limited to reach any definite
conclusion. It is an issue that is being studied further.

Generally speaking, one can look at our method and results also from a
purely condensed matter point of view, as an example of how the 
universal properties of spin and matter systems are affected by introducing
random elements into its definition.  
For example, the inclusion of random impurities in a lattice model leads to
Fisher renormalization \cite{fisher} of the critical exponents in the so-called 
annealed case, where the disorder forms part of the dynamics. 
By contrast, in the quenched case the Harris-Luck criterion
is conjectured to hold \cite{harrisluck}. 
To what extent these ideas extend to the case of random connectivity of the 
lattice -- which is the one relevant to non-perturbative quantum gravity models --
is still not very clear, despite a number of results in this direction, 
involving spin models with (quenched) geometric disorder 
coming from Euclidean dynamical triangulations and
Voronoi-Delaunay lattices (see, for example, \cite{jj,luck}).

One of the reference points for such investigations is that of the 
two-dimensional Ising model on 
fixed, regular lattices, which in absence of an external magnetic field can
be solved exactly in a variety of ways (see \cite{mccoywu,baxter}). In this
case, the critical exponents, which characterize the system's behaviour
near its critical temperature, are given by the so-called  
Onsager values $\alpha=0$, $\beta=0.125$ and $\gamma=1.75$ for the
specific heat, spontaneous magnetization and susceptibility. 
In sharp contrast, 
in the case of the Ising model on {\it Euclidean} dynamical triangulations 
mentioned earlier the disordering effect is so strong that the same
critical exponents are altered to $\alpha=-1$, $\beta=0.5$ and $\gamma=2$ 
\cite{kaz}. On the other hand, despite the
annealed geometric randomness present, the corresponding Ising model
coupled to {\it causal} dynamical triangulations seems to share
the flat-space exponents, indicating a perhaps surprising
robustness of the Onsager universality class.\footnote{The study of fermions
coupled to two-dimensional causal dynamical triangulations was initiated
in \cite{Burda}.}

In the present piece of work, we will concentrate for simplicity and
definiteness on the CDT-Ising system, although our method should also apply 
straightforwardly to other spin models  
and may inspire similar diagrammatic 
expansions in other, discrete models of quantum gravity. 
In the next section we will review the main features of causal dynamical
triangulations in two dimensions. Sec.\ 3 introduces 
the high-temperature expansion on regular and dynamical lattices. 
In Sec.\ 4, after recalling some general definitions and results of graph theory, 
we present a set of rules which in principle allow us to compute the weight
of any susceptibility graph on the fluctuating CDT lattice and its dual, and explain 
the method with the help of illustrative examples. The finite number of terms
of the series expansions we obtain in this way are analyzed in Sec.\ 5 with
the help of the ratio, Dlog Pad\'e, and differential approximants methods.
Sec.\ 6 contains some remarks on the extension of our method to 
low-temperature expansions, before we summarize our findings in Sec.\ 7.

\section{Causal Dynamical Triangulations revisited}

Despite the apparent failure of describing gravity as a renormalizable 
quantum field theory, the
hope that it may nevertheless admit a well-defined non-perturbative formulation has
been both inspiration and justification for several approaches to quantum gravity,
like Dynamical Triangulations \cite{DT-book}, 
Loop Quantum Gravity \cite{rovelli,lqg}, Exact Renormalization Group \cite{erge},
Spin Foams \cite{spinfoam} and Causal Sets \cite{cs}. 

The way to a non-perturbative quantization pursued by Dynamical Triangulation is that 
of giving meaning to the formal (Euclidean\footnote{An important feature of the {\it causal} 
version we will be using is that one 
starts with Lorentzian signature and a complex path integral
${\cal Z} = \int {\cal D} [g_{\mu\nu}] e^{i S_{EH}[ g_{\mu\nu}]}$, and then performs a Wick rotation
to the Euclidean version (\ref{pi}), which still retains a memory of the causal light-cone structure of
the original spacetimes. For the purposes of the current presentation we will 
work with the Euclideanized version of the Lorentzian CDT path integral throughout.})
path integral 
\beq \label{pi}
{\cal Z} = \int {\cal D} [g_{\mu\nu}] e^{- S_{EH}[ g_{\mu\nu}]}
\eeq
for the gravitational action, consisting of the usual Einstein-Hilbert term, a 
cosmological term, and possibly others.
To accomplish this task one starts from an initially discrete formulation where the sum 
over geometries,
i.e.\ the integral over metrics modulo diffeomorphisms, is defined as the continuum limit 
of a sum over
simplicial manifolds $T$. Roughly speaking, a simplicial manifold is a collection of 
$d$-dimensional triangular building blocks called ``simplices" (generalizing triangles
in $d=2$ and tetrahedra in $d=3$), glued together along their $(d-1)$-dimensional
faces (or edges in $d=2$) in such a way that local neighbourhoods look $d$-dimensional. 
To each simplicial manifold $T$ one can associate a Boltzmann weight $e^{-S_R}$, 
where the Regge action $S_R(T)$ represents a discrete version of the Euclidean 
gravity action. The simplicial building blocks
are taken to be all identical, with the length of each one-dimensional edge fixed to 
a cut-off value $a$. The thus regularized path integral takes the form
\beq
Z=\sum_{T}\frac{1}{C(T)}e^{-S_R(T)},
\eeq
where $C(T)$ is a symmetry factor. Note that the diffeomorphism invariance
of the continuum formulation has been taken care of by adopting
an explicitly 
coordinate-independent formulation. (We refer to \cite{ajl1} for details of motivation and 
construction.)

This approach requires two restrictions on the ensemble of simplicial manifolds to
make it a viable candidate for a four-dimensional theory of quantum gravity. 
First, the spacetime topology must be fixed, since otherwise the 
number of inequivalent geometries grows factorially with the volume and the Boltzmann 
weight would have no chance of controlling the resulting divergence.
The second restriction has emerged only after a detailed study of the dynamics of 
the dynamically triangulated model. It turned out that the original, purely Euclidean
model behaved too wildly to admit a sensible classical limit. In particular, no region in
the phase space of the underlying statistical model of random geometry was found
where the large-scale geometry was extended and four-dimensional (as, for example,
measured by its large-scale Hausdorff dimension $d_H$). 

The restriction proposed in \cite{al} to resolve this impasse was 
to impose a causal structure on the 
triangulated spacetimes in such a way that the simplices
interpolate between successive spatial slices\footnote{A slice is defined as 
a $(d-1)$-dimensional spatial submanifold
of constant topology $\Sigma$ of the simplicial spacetime manifold. 
This fixes the spacetime topology to be of type
$\Sigma\times \mathbb{R}$. For definiteness, we will use $\Sigma=S^1$ in
what follows.} of constant integer proper time.
The ensemble of geometries with these two restrictions implemented underlies
the approach of {\it Causal} Dynamical Triangulations. So far, CDT have lived up
to their original expectation: to the best of current knowledge, the quantum 
geometries generated dynamically
from simplicial building blocks of dimension $d=2,3$ and 4 have an effective 
large-scale dimension equal to $d$ \cite{al,3dcdt,ajl-prl,spectral}. 
 
\begin{figure}[ht]
\centering
\vspace*{13pt}
\includegraphics[width=11cm]{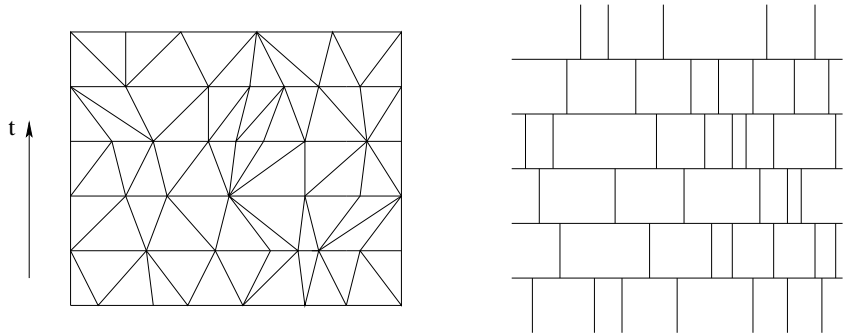}
\vspace*{13pt}
\caption{\footnotesize Example of a triangulated piece of spacetime in the CDT model 
(left) and the corresponding dual graph (right).
The picture captures the way triangles are glued together, but does not represent
faithfully intrinsic distances, and therefore the curvature properties of the simplicial 
geometry. If we were to respect this property we would not be able to draw the picture
in a plane, because the model's triangles are all equal and equilateral.}
\label{cdt}
\end{figure}

In this paper we shall restrict ourselves to the case $d=2$, which brings about several
simplifications for the CDT model, including a graphical one, in the sense that
triangulated
spacetimes from the causal ensemble just introduced can be represented by
planar drawings, like the one in Fig.\ \ref{cdt}.
In the same figure we also show the corresponding dual graph, where triangles 
have been substituted by trivalent vertices
and the gluing of neighbouring triangles by links connecting the corresponding 
vertices. The representation by dual graphs has been used in some analytical
investigations \cite{difra}, and will be used in the present article when we
discuss the ``dual" Ising model with the spins placed on triangles instead of
vertices.

A second and more important simplification in the two-dimensional case is that 
the Einstein-Hilbert term (the integrated scalar curvature) is purely topological
and can therefore be dropped from the action, leaving only the cosmological
term proportional to the volume of the manifold. This means that in our
approach in $d=2$ the path integral (\ref{pi}) is replaced by the partition function
\beq \label{pure}
Z=\sum_{N} e^{-\lambda N} \sum_{T_{N}} \frac{1}{C_{T_{N}}},
\eeq
where $T_{N}$ labels distinct (causal) triangulations of fixed volume $N$, the number 
of triangles\footnote{With all triangles equilateral of side $a$, the volume of each triangle is
$\frac{\sqrt{5}}{4}a^2$, and hence the total volume proportional to $N$.}, and $\lambda$ is
the bare cosmological constant.
Asymptotically (\ref{pure}) can be shown to behave like
\beq
Z=\sum_{N} e^{(\lambda_c-\lambda)N+o(N)},
\eeq
with $\lambda_c=\log 2$. It is straightforward to see that the model is well 
defined for $\lambda>\lambda_c$ and that in the
limit of $\lambda\rightarrow\lambda_c$ the average discrete volume $\langle N\rangle$ 
goes to infinity.
In this limit one can send the cut-off $a$ to zero while keeping the physical volume
$V=\frac{\sqrt{5}}{4}a^2 N$ finite, thus obtaining the continuum limit of the model.

\section{The Ising model and its high-$T$ expansion}

Given any lattice $G$ of volume $N$ with $v$ vertices and $l$ links (edges), 
the Ising model on $G$ in the presence of
an external magnetic field $h$ is defined by the partition function
\beq
Z_N(K,H)=\sum_{\{\sigma_i=\pm 1\}_{i\in G}}e^{-\beta \mathcal H[\{\sigma\}]}
=\sum_{\{\sigma_i=\pm 1\}_{i\in G}}e^{K \sum_{\langle i j \rangle}\sigma_i\sigma_j +H\sum_i\sigma_i},
\eeq
where $ \mathcal H[\{\sigma\}]$ is the Hamiltonian of the spin system, $\beta:=1/(k_B T)$
the inverse temperature,
$\sigma_i$ the spin variable at vertex $i$, taking values $\sigma_i=\pm 1$, and
${\langle i j \rangle}$ denotes nearest neighbours. In standard notation, 
we will use $K=\beta J$, where $J>0$ is the ferromagnetic spin coupling, and 
$H=\beta mh$, where $m$ is the spin magnetic moment.
The fact that $\sigma_i\sigma_j=\pm1$ allows us to write
\beq
\begin{split}
e^{K\sigma_i\sigma_j} &=(1+u\; \sigma_i\sigma_j) \cosh (K),\\
e^{H\sigma_i} &=(1+\tau\; \sigma_i) \cosh (H),
\end{split}
\eeq
in terms of newly defined variables $u=\tanh (K)$ and $\tau=\tanh(H)$,
and rewrite the partition function as 
\beq\label{high-exp}
\begin{split}
Z_N(K,H)
   &=\sum_{\{\sigma_i\}}\prod_{\langle i j \rangle}e^{K\sigma_i\sigma_j}\prod_i e^{H\sigma_i}\\
   &=\cosh^l(K) \cosh^v(H)\sum_{\{\sigma_i\}}\prod_{\langle i j \rangle}(1+u\; \sigma_i\sigma_j)
     \prod_i (1+\tau\; \sigma_i)\\
   &=\cosh^l(K) \cosh^v(H)\sum_{\{\sigma_i\}}\big[ (1+u\sum_{\langle i j \rangle}\sigma_i\sigma_j+
     u^2\sum_{\langle i j \rangle,\langle k l \rangle,j\neq l}\sigma_i\sigma_j\sigma_k\sigma_l+...)\times\\
   & \hspace{1.5cm}\times (1+\tau \sum_i\sigma_i+\tau^2\sum_{i\neq j}\sigma_i\sigma_j+...) \big]\\
   &=2^v \cosh^l(K) \cosh^v(H)\; F^{(N)}(u,\tau).
\end{split}
\eeq
Because of the sum over spin values $\pm 1$ it is clear that every term containing a spin at a 
given vertex to some odd power will give a vanishing contribution. Thus we need to keep only 
terms with even powers of $\sigma_i$'s, which are terms where each $\sigma_i$ belongs 
either to an even number of nearest neighbour couples $\sigma_i\sigma_j$ (from the 
$u$-terms) or to an odd number of them and to one of the spins coming from the 
$\tau$-terms. It is not difficult to convince oneself that each such term
is in one-to-one correspondence with a graph drawn on the lattice whose length is 
given by the power of $u$
and whose number of vertices with odd valence is given by the power of $\tau$.
The function $F^{(N)}(u,\tau)$ introduced in eq.\ (\ref{high-exp}) is then the generating 
function for the number of such graphs that can be drawn on the lattice, and the factor 
$2^v$ comes from the sum over the spin configurations.
The representation we have obtained in this manner is a high-temperature 
expansion, since for infinite temperature $T$ we have $u=\tau=0$. 
We will return to this graphic interpretation in the next section, after having recalled 
some basic definitions from graph theory. In what
follows, we will only be interested in the case of vanishing external magnetic field.

All we have said up to now does not require any specific properties of the lattice 
$G$, but works for any lattice, regular or not. This makes it
straightforward in the framework of CDT to couple the Ising model to gravity. 
We simply associate a complete Ising model with each
triangulation of the ensemble, viewed as a lattice, by putting the spins at 
the vertices of the triangles,
and then perform the sum over such triangulations, leading to
\beq
\tilde{Z}(K,H,N)=\sum_{T_N}\;\sum_{\{\sigma_i\}_{i\in T_N}}e^{-\beta \mathcal H[\{\sigma\}]}.
\label{sumtri}
\eeq
The situation where the volume is allowed to fluctuate is also of interest in 
a quantum gravity context and described by the grand canonical partition function
\beq
\tilde{\mathcal Z}(K,H)=\sum_N e^{-\lambda N}\tilde{Z}(K,H,N),
\eeq
where the role of chemical potential is played by the cosmological constant $\lambda$.
As usual in dynamically triangulated models, the asymptotic behaviour as function of
$N$ of the canonical partition function in the infinite-volume limit $N\rightarrow\infty$ 
will determine the critical value of $\lambda$ at which the continuum limit can be performed.

In trying to understand the behaviour of the Ising model coupled to quantum
gravity in the form of causal dynamical triangulation, we will make crucial use
of the known probability distribution $P(q)$ of the coordination number $q$ of 
vertices (the number $q$ of links meeting at a vertex) for the {\it pure} gravity case. 
This is relevant because we will be expanding about the point $\beta =0$, at
which the spin partition function (\ref{high-exp}) reduces to a trivial term $2^v$ and
the geometry is therefore that of the pure 2d Lorentzian gravity model. 
The probability distribution in the thermodynamic limit was derived in \cite{aal1}, 
resulting in
\beq
P(q)=\frac{q-3}{2^{q-2}}.
\label{vertprob}
\eeq
\begin{figure}[ht]
\centering
\vspace*{13pt}
\includegraphics[width=11cm]{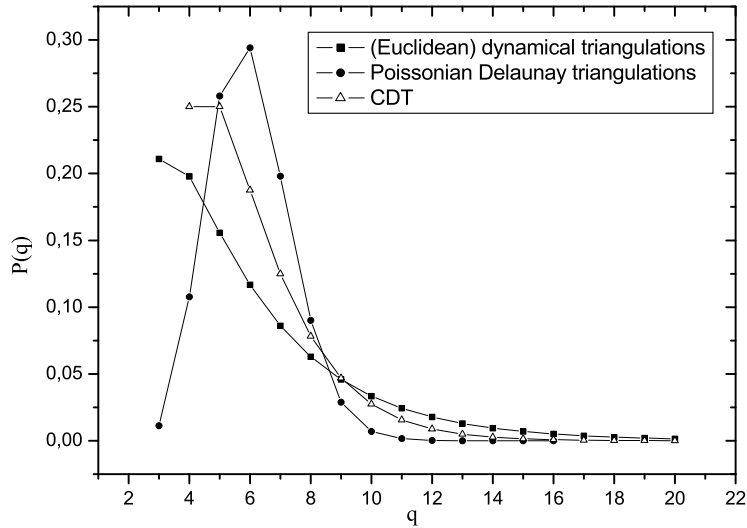}
\vspace*{13pt}
\caption{\footnotesize Comparing the probability distributions $P(q)$ of the
vertex coordination numbers $q$ for three different types of random triangulations:
Euclidean dynamical triangulations \cite{randomedt}, Poissonian Delaunay triangulations
\cite{drouffe} and causal dynamical triangulations \cite{aal1}.}
\label{coord-distr}
\end{figure}
The main idea is to use this distribution, together with some information about correlations
between distributions at different vertices, to re-express the sum over triangulations
in the high-temperature expansion of the partition function, (\ref{sumtri}), by an
average over coordination numbers. In terms of geometric randomness, the pure 
CDT model lies in between the Voronoi-Delaunay triangulations
based on Poissonian random distributions of vertices and the planar
triangulations underlying the approach of
Euclidean dynamical triangulations. A comparative plot of the probability distributions
for the coordination numbers in the three different types of geometry
is shown in Fig.\ \ref{coord-distr}. 

Let us point out a further subtlety which arises from the fact that the underlying 
lattices are not fixed, but fluctuating, and therefore do not have a fixed ``shape".
In this case it can happen that in the graph counting for the high-temperature
expansion of some thermodynamic quantity topologically non-trivial graphs must
be taken into account. By this we mean closed non-contractible graphs which wind 
around the
spacetime. This does not invalidate the method in
principle, but requires a more detailed knowledge of the global properties of the
spacetime. In two-dimensional CDT, where the spacetime topology is usually
chosen to be that of a cylinder (a spatial circle moving in time), this would be
closed loop graphs which wind one or more times around the spatial $S^1$, and 
might be as short as a single link. Such pinching configurations certainly exist,
but their contribution to the graph counting has been shown to be irrelevant in
CDT, because it is subleading in $N$ in the thermodynamic limit \cite{aal1}.

\section{Graph embeddings}

\subsection{Terminology of graph theory and embeddings}

To prepare the ground for our counting prescription, we will first
review some of the relevant terminology of graph theory and embeddings,
along the lines of reference \cite{domb1}.

A linear graph is a collection of $v$ vertices and $l$ lines (or edges) 
connecting pairs of vertices.
A {\it simple graph} is a graph in which two vertices are connected by at most one line 
and in which single lines are not allowed to form closed loops.
A graph is said to be {\it connected} if there is at least one path of lines between any two 
vertices, and {\it disconnected} if for some vertex pair there is no such path.

The {\it cyclomatic number} $c$ of a connected graph $g$ is defined as
\beq
c(g)=l-v+1,
\eeq
and represents the number of independent cycles in the graph.

The {\it degree} (or {\it valence} or {\it coordination number}) of a vertex is the number 
of edges incident on that vertex.

Two graphs are said to be {\it isomorphic} if they can be put into one-to-one 
correspondence in such a way that their vertices and edges correspond. They are
called {\it homeomorphic} if they are isomorphic after insertion or suppression of any 
number of vertices of degree 2 (this operation being defined on an edge in a trivial way).

\begin{figure}[ht]
\centering
\vspace*{13pt}
\includegraphics[width=12cm]{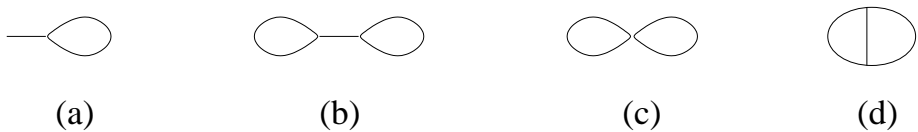}
\vspace*{13pt}
\caption{\footnotesize Four kinds of homeomorphs: (a) tadpole, (b) dumbbell, (c) figure eight,
and (d) $\theta$-graph.}
\label{topol}
\end{figure}

Homeomorphs are thus graphs with the same topology, in particular, with the same 
cyclomatic number.
It is then possible to classify graphs in terms of {\it irreducible} graphs. An irreducible 
graph is one in which all vertices of degree 2 have been suppressed. 
(Note that in general it will no longer be a simple
graph since it will contain single-line loops.)
Typical irreducible graphs whose homeomorphs we will encounter are 
{\it tadpoles, dumbbells, figure eights}
and {\it $\theta$-graphs} (Fig.\ \ref{topol}).

In the high-temperature expansion of an Ising model it is useful to think of the lattice 
with its $v$ vertices and $l$ edges as a graph $G$. The graphs appearing in the 
expansion of the thermodynamic function of interest are then graphs embedded in $G$.
A general definition of embeddings of graphs is the following.
Let $G$ be a general graph. A graph $H$ is a {\it subgraph} of $G$ when all its vertices 
and edges are vertices and edges of $G$. An {\it embedding}\footnote{To be precise, 
this defines a {\it weak} embedding. By contrast,
a {\it strong} embedding of $g$ in $G$ is defined as any section
graph $G^+$ of $G$ which is isomorphic to $g$, where a section
graph is a subgraph of $G$ consisting of a subset $A$ of vertices
and all the edges which connect pairs of (nearest-neighbour) vertices
of $A$. Since all our calculations will concern weak embeddings and the
corresponding weak lattice constants,
we will from now on drop this specification.} 
of a graph $g$ in $G$ is a subgraph $G'$ of $G$ which
is isomorphic to $g$.

The {\it lattice constant} $(g;G)^{(N)}$ of a graph $g$ on $G$ is defined as the number 
of different embeddings of $g$ in $G$. In practice it is often useful to work with
the lattice 
constant {\it per site}, defined as $(g;G)=\frac{2}{N}(g;G)^{(N)}$.

With these definitions in hand we can now rewrite the function $F^{(N)}(u,\tau)$ 
introduced in (\ref{high-exp}) as
\beq
F^{(N)}(u,\tau)\equiv 1+\sum_{s=0}^{\infty}\sum_{l>0} D_{l,2s}^{(N)}u^l\tau^{2s},
\eeq
where $D_{l,2s}^{(N)}$ are the lattice constants of graphs of length $l$ and with $2s$ odd 
vertices, for which $F^{(N)}(u,\tau)$ serves as a generating function.
It follows from the extensive nature of the free energy of the system that
in the thermodynamic limit 
\beq
F^{(N)}(u,\tau)\stackrel{N\rightarrow\infty}{ \longrightarrow} {\rm e}^{N\Theta (u,\tau)}=
1+ N\Theta(u,\tau) +O(N^2),
\label{non}
\eeq
where the function $\Theta (u,\tau)$ does not depend on $N$.
From the usual combinatorial theory of non-embedded graphs it is well known that the 
logarithm of the generating function is the generating function for connected graphs. 
Because of the embedded nature of the graphs in the case at hand,
the disconnected graphs will still contribute to $\log F^{(N)}(u,\tau)$ with a 
``repulsion" term of sign $(-1)^{(\# connected\ components)-1}$. 

In our analysis of the Ising model on CDT we will be looking at a quantum-gravitational
average of the magnetic susceptibility at zero external field. For the
usual Ising model the susceptibility $\chi$ per unit volume is given by
\beq\label{chi}
   \chi(u)=\frac{1}{N}\frac{\partial^2\ln Z_N}{\partial H^2}_{\big|H=0}=
   \frac{1}{2}+\frac{2}{N}\frac{\sum_l D_{l,2}^{(N)}u^l}{1+\sum_l D_{l,0}^{(N)}u^l},
\eeq
where in the second step we have substituted in the high-temperature expansion
of the partition function. By virtue of (\ref{non}) it is clear that the susceptibility is
independent of $N$ in the infinite-volume limit (thus justifying our notation
$\chi(u)$), and that moreover the denominator in the last term in (\ref{chi})
does not contribute at lowest order in $N$, which is the one relevant
to the computation. To calculate the susceptibility, it is therefore sufficient
to compute the term of order $N$ in $\sum_l D_{l,2}^{(N)}u^l$, as is well known.

The computation in the gravity-coupled case proceeds completely analogously.
Performing the sum over all triangulations at fixed $N$, as in eq.\ (\ref{sumtri}),
and then letting $N\rightarrow\infty$, one obtains 
\beq\label{chigrav}
   \chi_{\rm CDT}(u)\approx 
   \frac{2}{N}\, \frac{\sum_l\left( \sum_{T_N}D_{l,2}^{(N)}(T_N)\right) u^l}
   {\sum_{T_N}1+\sum_l \left( \sum_{T_N}D_{l,0}^{(N)}(T_N)\right) u^l}=
    \frac{2}{N}\, \frac{\sum_l\langle D_{l,2}^{(N)}\rangle u^l}
   {1+\sum_l \langle D_{l,0}^{(N)}\rangle u^l}
\eeq
for the susceptibility in 
presence of the gravitational CDT ensemble, where we have dropped the irrelevant
constant term and introduced an obvious notation for the ensemble average in
the second step. Whereas it was fairly straightforward to verify the cancellations
of terms of higher order in $N$ between numerator and denominator in the pure
Ising case in eq.\ (\ref{chi}) by an explicit calculation, the analogous computation
does not seem feasible in the gravity-coupled case, although we know from
the same general arguments that it must be realized here too. Again the terms
in the denominator of the expressions on the right-hand side of (\ref{chigrav}) do
not contribute at lowest order, which -- at least for this particular observable -- 
implies that its annealed and quenched gravitational averages coincide.

\subsection{Useful results on lattice constants}

We now want to identify which graph topologies appear in the 
high-temperature expansion of the magnetic susceptibility.
For convenience we can divide our set of graphs into two subsets,
graphs with zero cyclomatic number (also called Cayley trees), and graphs
with $c>0$. Since the graphs we encounter in the susceptibility
series must have two and only two odd vertices, it follows that the first subset will 
only contain connected non-selfintersecting open chains (in other words, self-avoiding 
walks), while the second will contain everything else.
The connected graphs in the second subset are graphs with loops and one or no open ends,
while the disconnected ones take the form of a union of one of the previous two kinds 
with any number of even graphs (graphs in which every vertex has even degree).

The problem we meet in the high-temperature expansion is the evaluation of the 
lattice constants of such graphs. It turns out that there are two kinds of theorems 
which can be applied profitably in the context of our random triangulated lattices.
The first is very general and reduces the calculation of the lattice constants for 
disconnected graphs into that for connected graphs. This makes it possible to 
compute the lattice constants for the second subset ($c>0$) in a unified way.
The second theorem is less general but powerful, since it gives us a recurrence 
relation for the expansion coefficients of the susceptibility.

\begin{figure}[ht]
\centering
\vspace*{13pt}
\includegraphics[width=13cm]{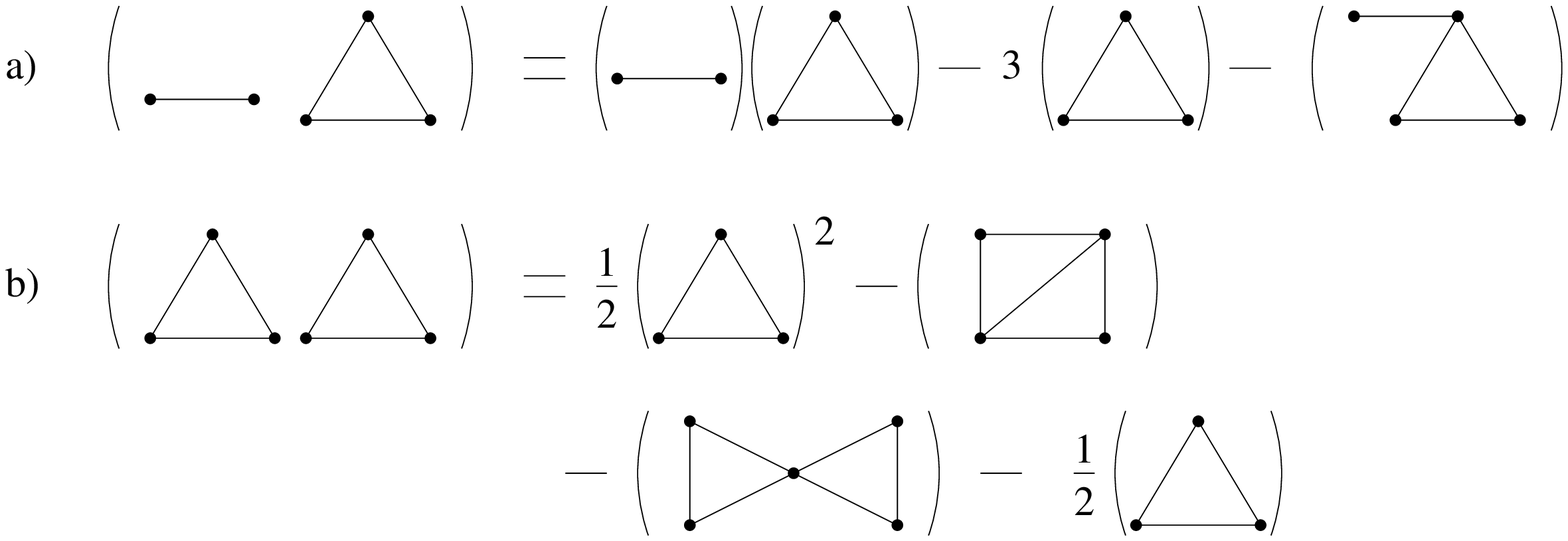}
\vspace*{13pt}
\caption{\footnotesize Two examples of the reduction theorem. The graph $G$ 
has been dropped in the notation.}
\label{reduct}
\end{figure}

The {\it reduction theorem} for disconnected graphs states that if $g_i$ and $g_j$ 
are two graphs $g_i\neq g_j$ and $G$ is any graph, then
\beq \label{reduction}
(g_i\cup g_j;G)=(g_i;G)(g_j;G)-\sum_k\{ g_i+g_j=g_k\}(g_k;G),
\eeq
where $g_i\cup g_j$ stands for the disjoint union of the two graphs, 
$\{ g_i+g_j=g_k\}$ is the number of possible choices
of embeddings of $g_i$ and $g_j$ in $g_k$ having $g_k$ as their sum graph, and the 
summation is over all graphs $g_k$ obtainable in this way.
If $g_i= g_j$ we can compute the right-hand side of (\ref{reduction}) as if the two graphs 
had different colours
and thus obtain $2 (g_i\cup g_i;G)$. If $g_i$ or $g_j$ are themselves 
non-connected graphs we can iterate
the theorem. In this way the lattice constant of the disconnected union of $n$ connected 
graphs can be expressed
as a polynomial of order $n$ in the lattice constants of connected graphs (see Fig.\ \ref{reduct}
for some simple examples).

The {\it counting theorem} for the susceptibility series was first given by Sykes \cite{sykes} in 1961,
but only later proved by Nagle and Temperley \cite{nagletemp} as a special case of a more
general theorem of graph combinatorics. Closer examination reveals that the proof 
makes no reference to the structure of the lattice, but only to its coordination number $q$,
which will turn out to be useful when considering the lattice dual to the triangulation.
The theorem states that the high-temperature series for the susceptibility can be written as
\beq
\chi(u)=1+(1-\sigma u)^{-2}\left[ q u (1-\sigma u)-2 (1-u^2) S_1 + 8(1+u)^2(S_2+S_3)\right],
\label{expa}
\eeq
where $q=\sigma +1$ is the coordination number, and the functions $S_i$ are defined as
\beq
\begin{split}
S_1 &= \sum_{l>0} l d_l u^l,\\
S_2 &= \sum_{l>0}\left[\sum_{w>0} w d_l(w)\right] u^l ,\\
S_3 &=\sum_{l>0} \left[\sum_{r\leq r'} rr'm_l(r,r')\right] u^l.
\end{split}
\eeq
The coefficient $d_l$ is the sum of lattice constants of even graphs with $l$ lines, 
while $d_l(v)$
is restricted to even graphs with $l$ lines and vertices of degree $2q_1, 2q_2,...$, 
characterized by the
weight $w=\sum_i q_i(q_i-1)/2$.
The function $m_l(r,r')$ is the sum of lattice constants of graphs with exactly two odd 
vertices of degrees $2r+1$
and $2r'+1$.
Substituting the left-hand side of formula (\ref{expa}) by the expansion 
$\chi (u)= 1+\sum_{n>0} a_n u^n$, one finds
a recursive relation for the coefficients $a_n$, namely,
\beq \label{recurr}
\begin{split}
a_l-2\sigma a_{l-1} &+\sigma^2 a_{l-2} = 2(l-2)d_{l-2}-2ld_l+ \\
&+8\left[\sum_{w>0} w d_l(w)+2\sum_{w>0} w d_{l-1}(w)
+\sum_{w>0} w d_{l-2}(w)\right]+\\
&+8\left[\sum_{r\leq r'} rr'm_l(r,r')+2\sum_{r\leq r'} rr'm_{l-1}(r,r')+
\sum_{r\leq r'} rr'm_{l-2}(r,r')\right].
\end{split}
\eeq
As a result, at every new order $l$ in our calculation the only new lattice constants we 
have to evaluate are those
for graphs with no vertex of degree 1, reducing the computational effort considerably.

This formula can in principle be applied to any non-regular lattice whose vertices are 
all of the same degree. For a fixed non-regular lattice this observation would usually 
not be of much help, because one would still need to compute at each order the new
lattice constants depending on the complete, detailed information of the lattice geometry.
In our case this difficulty is not present, since we are summing over triangulations
and can simply substitute in (\ref{recurr}) the
lattice constants {\it averaged} over the triangulations, without the need of keeping
track of the lattice geometry for individual lattices.

\subsection{Counting graphs on the CDT triangular lattice}

Next, we will analyze the evaluation procedure for lattice constants in the 
quantum-gravita\-tio\-nal CDT model.
Computations here are made difficult by the randomness of the coordination 
number.
At the outset it is not even obvious that any of the known recurrence relations can be used.
Our task is to list all possible diagrams, and for each count the number of ways
it can be embedded in the lattice. On a regular lattice such an operation is tedious
but straightforward: starting from a fixed vertex, we trace out all possible sequences 
of links of a given length, say, which do not self-intersect. 
The discrete symmetries of the regular lattice simplify this task greatly.
In the simplest case, the counting has to be done only for a single initial vertex, with
all other graphs obtained subsequently by translational symmetry.  
The analogous counting on the dynamical CDT lattices 
is complicated by the fact that the lattice 
neighbourhoods of different vertices will in general look different. We will deal
with this difficulty by setting up an algorithm to count the embedding of a given
diagram on the {\it ensemble} of CDT lattices, making use of the known
probability distribution of the vertex coordination numbers. 

\begin{figure}[ht]
\centering
\vspace*{13pt}
\includegraphics[width=8cm]{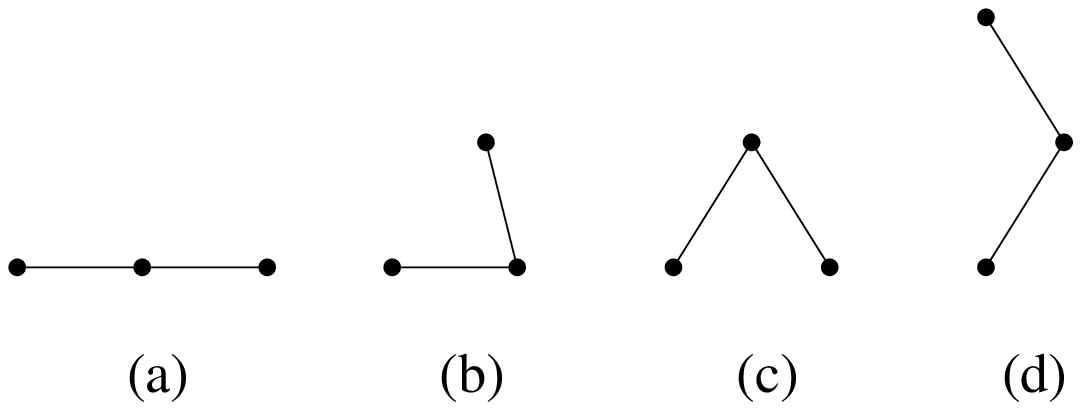}
\vspace*{13pt}
\caption{\footnotesize The possible typologies of embedding, up to reflection symmetry, 
of a length-2 open chain.
The angle which the time-like links form with the vertical has no relevance for 
the graph counting.}
\label{embed2}
\end{figure}

\begin{figure}[ht]
\centering
\vspace*{13pt}
\includegraphics[width=8cm]{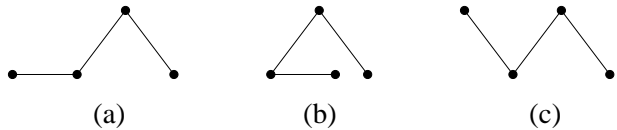}
\vspace*{13pt}
\caption{\footnotesize Three of the different typologies of possible embeddings of 
a length-3 open chain.}
\label{embed3}
\end{figure}

Before presenting this algorithm we need some more notation.
CDT lattices have two kinds of links, time-like and space-like\footnote{
Even after performing the Wick rotation to Euclidean signature, these two
link types remain combinatorially distinguishable because of the special way
in which the original Lorentzian simplicial geometries were constructed.}.
In keeping with the usual representation of two-dimensional CDT lattices (c.f.
Fig.\ \ref{cdt}, left-hand side), we will draw time- and space-like links in embeddings 
of graphs as (diagonally) upward-pointing and horizontal lines, respectively 
(see Fig.\ \ref{embed2}). Furthermore we need to keep track of the relative
up-down or right-left orientation of consecutive links, to distinguish between
graphs like (c) and (d) in Fig.\ \ref{embed2} (which give
different contributions to the lattice constant). 
In addition, the counting for a graph like (b) will be identical to that for its
mirror images under left-right and up-down reflections. These mirrored graphs will
be counted by multiplying the embedding constant of the graph by an 
appropriate {\it symmetry factor}.
A symmetry factor 1 is assigned to graphs symmetric under both up-down
and left-right reflections (for example, graphs $(a)$ and $(d)$), a factor 2 to graphs 
with only one of the two symmetries (like graph $(c)$) or symmetric with respect
to a composition of the two
(like graph $(c)$ of Fig.\ \ref{embed3}), and a factor 4 to graphs without
reflection symmetry (like graph $(b)$
of Fig.\ \ref{embed2} or $(a)$ and $(b)$ of Fig.\ \ref{embed3}).

The lattice constant $(g;T)$ of a graph $g$ embedded in a graph corresponding 
to a triangulation $T$ is computed as
\beq\label{lat-cnst}
(g;T)=\sum_i s_i (g;T)_i,
\eeq
where $(g;T)_i$ is the number of ways a particular typology of embedding $(g)_i$ 
can be realized on $T$,
$s_i$ is the relevant symmetry factor, and the sum extends over all possible 
typologies of embeddings of the graph $g$.
In the context of quantum gravity, we are interested in the average of
(\ref{lat-cnst}) over all triangulations, that is,
\beq
\langle g\rangle = \frac{\sum_T(g;T)}{\sum_T 1},
\eeq
which will take into account the 
probability distribution of the vertex coordination.

As mentioned earlier, all calculations will be performed in the thermodynamic limit 
where the cosmological constant
is tuned to its critical value $\lambda_c=\ln 2$. 
In this limit, the probability of having $k$ incoming
time-like links at a particular lattice vertex is given by
\beq \label{prob}
p(k)=\frac{1}{2^k},
\eeq
with an identical probability for having $k$ {\it outgoing} time-like links at a vertex.
Moreover, {\it at one and the same vertex} the two probabilities are independent of each 
other. In fact, not just for a single vertex are these two probabilities independent,
but the same is true for all vertices lying in the same space-like, horizontal slice.
By contrast, the outgoing probability at a vertex will in general condition the incoming 
probability at another vertex on the subsequent horizontal slice. By construction,
the probability of having a space-like link to the left and to the right of an given
vertex is equal to 1.

Armed with this information we are now ready to compute any of the $(g;G)_i$ in 
(\ref{lat-cnst}). We do not have a general counting formula in closed form, but
will formulate a number of rules and tools which will enable us to
do the counting recursively.
We will start with some illustrative examples. Examples 3 and 4 will serve as 
our elementary building blocks in more complicated constructions.\\

\noindent\textbf{Example 1.} The simplest example is that of a length-1 graph $c_1$, 
consisting of
a single horizontal or vertical link. The number of horizontal embeddings
is the number of horizontal links on a CDT lattice, which by virtue of the
Euler relation is given by half the number of triangles, i.e. $N/2$.
If we give the lattice constant per vertex, as is customary in the literature,
it is therefore $\langle c_1\rangle_h=1$.
By a similar topological argument we can immediately deduce the number
2 for the lattice constant of vertical embeddings, but it is instructive to compute it
from the probability distribution (\ref{prob}) instead. If at a vertex we have $k$ 
outgoing vertical links, there are precisely
$k$ ways to embed the one-link graph such that it emanates in upward direction from
this vertex. Since every vertical embedding is accounted for in this way, by 
assigning the relevant probability and summing over
$k$ we obtain
\beq
\langle c_1\rangle_v=\sum_{k=1}^{\infty}\frac{k}{2^k}=2,
\eeq
in agreement with the earlier argument. The total contribution to the
lattice constant at order 1 is therefore
$\langle c_1\rangle_h+\langle c_1\rangle_v=3$.\\

\noindent\textbf{Example 2.} Next, let $c_2$ be a length-2 open chain, and consider the embedding 
typology $c_{2,b}$ shown in Fig.\ \ref{embed2}$(b)$. 
The result follows straightforwardly from example 1: there is one horizontal link per vertex,
and there are in the ensemble on average two ways of attaching an outgoing vertical
line to its right vertex, yielding $\langle c_2\rangle_b=2$.
For the embedding of Fig.\ \ref{embed2}$(a)$ there is nothing to compute; we have 
$\langle c_2\rangle_a=1$. The embedding
of Fig.\ \ref{embed2}$(d)$ is the first configuration we encounter which extends over 
two triangulated strips. Because of the independence
of the probabilities (\ref{prob}) in successive strips we obtain
\beq
\langle c_2\rangle_d= (\sum_{k=1}^{\infty}\frac{k}{2^k})^2=4.
\eeq
\\
\noindent\textbf{Example 3.} Now consider the embedding $c_{2,c}$ depicted in 
Fig.\ \ref{embed2}$(c)$. If the vertex on top
has $k$ incoming links, there are exactly $k(k-1)/2$ ways to realize the embedding,
leading to
\beq
\langle c_2\rangle_c=\sum_{k=1}^{\infty}\frac{k (k-1)}{2^{k+1}}=2.
\eeq
Note that the same ``inverted-V" embedding on a flat triangular lattice
$T_{\rm reg}$ would give $(c_2;T_{\rm reg})_c=1$ instead.
Putting everything together and multiplying by the appropriate symmetry 
factors we get 
$\langle c_2\rangle_a+4\langle c_2\rangle_b+2\langle c_2\rangle_c+\langle c_2\rangle_d=17$
as a total contribution at order 2.

\begin{figure}[ht]
\centering
\vspace*{13pt}
\includegraphics[width=10cm]{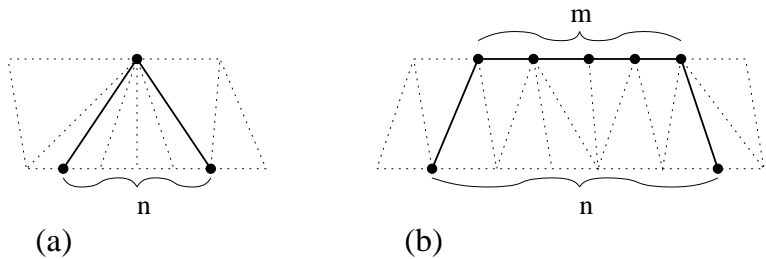}
\vspace*{13pt}
\caption{\footnotesize Two configurations contributing to the calculations of 
examples 3 and 4. The labels m and n count the numbers of space-like links as
indicated.}
\label{kernel}
\end{figure}

In view of more complicated graphs, an alternative and more convenient way 
of doing the calculation in example 3 is by focussing
on the probability distributions of the vertices lying on the {\it lower} space-like slice 
of the triangulated strip (see also Fig.\ \ref{kernel}$(a)$).
The probability of having $n$ space-like links in between the two vertices 
at the ends of the inverted V is given by 
the probability that there are $n-1$ vertices in between with precisely one outgoing 
link each, which is $1/2^{n-1}$.
Thus we obtain again
\beq
(c_2;G)_c=2\sum_{n=1}^{\infty}\frac{1}{2^{n-1}}=4.
\eeq
\\
\noindent\textbf{Example 4.} Consider an open chain of length greater than 
2 embedded in such a way
that only the first and last edges lie along time-like links, with a horizontal chain of $m$ space-like 
links in between. Both time-like links are supposed to lie in the same strip of the triangulation,
as illustrated in Fig.\ \ref{kernel}$(b)$. We are going to prove that the lattice constant of
such a configuration is given by
\beq\label{c_mn}
\langle c_{m+2}\rangle_b=\sum_{n=1}^{\infty}\frac{1}{2^{m+n-1}}\matrice m+n \\ m 
\ematrice = 4- 2^{1-m}.
\eeq
The result $4-2^{1-m}$ could be obtained easily by (i) considering
the number of ways in which each of the two vertices at the top left and right can have an
incoming link (giving a total of $2\times 2=4$ possibilities) and (ii) subtracting from that
the probability for the two time-like links touching each other in their lower vertex, 
which is $2^{1-m}$ (the probability that the $m-1$ intervening vertices have only one 
incoming link each). However, we rather want to do the
counting in a way that keeps track of the number of links separating the two vertices
at the bottom, for reasons that will become clear soon.
In other words, we want to determine the lattice constant for a closed polygon 
consisting of two
vertical edges, $m$ edges along the top and $n$ along the bottom.
From a combinatorial point of view this can be rephrased as the problem of counting 
the number of different ways in which $n$ triangles and $m$ upside-down triangles can 
be arranged to form a strip, with the well-known
binomial result $\big({m+n\atop m}\big)$.
To get the lattice constant for our random lattice we still have to include a probabilistic
factor $\frac{1}{2}$ for each vertical link in the polygon interior, leading to the factors 
$2^{1-m-n}$ in formula (\ref{c_mn}).\\

\begin{figure}[ht]
\centering
\vspace*{13pt}
\includegraphics[width=10cm]{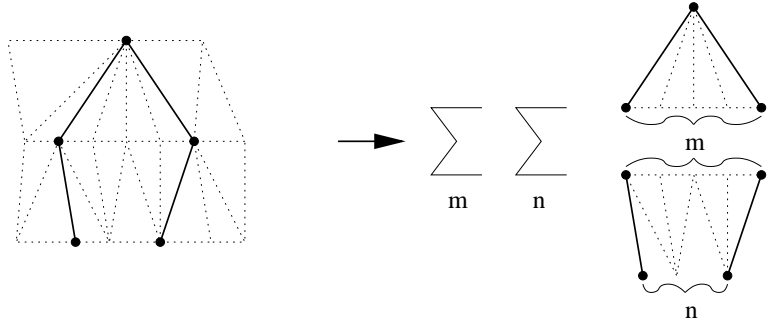}
\vspace*{13pt}
\caption{\footnotesize Example of a strip decomposition. The embedding on the left is split
into two strips, after which the strip contributions for given $m$ are multiplied pairwise
and summed over.}
\label{convol}
\end{figure}

\begin{figure}[ht]
\centering
\vspace*{13pt}
\includegraphics[width=10cm]{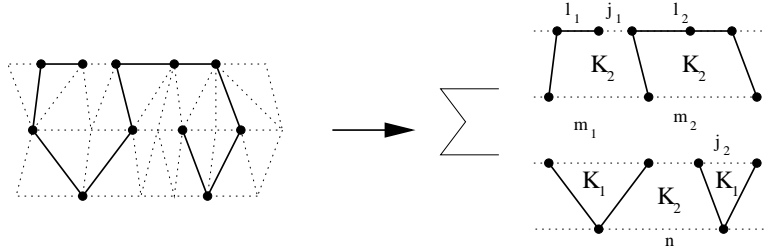}
\vspace*{13pt}
\caption{\footnotesize Example of a more complicated strip decomposition where each strip
consists of a sequence of elements of the kind illustrated in Fig.\ \ref{kernel}. 
The sum refers to the total sum in the expression (\ref{ex-stripdec})
which has to be performed after the proper weights have been assigned
to each element of the graph.}
\label{convol2}
\end{figure}

\noindent\textbf{The strips decomposition.}
The embedding typology $(g)_i$ of any connected graph $g$ extends over a 
well-defined number of strips in the CDT triangulation, and each part of $(g)_i$
belongs to a definite strip or horizontal slice.
Our first step in the counting procedure is to decompose the embedded graphs
into strip contributions, thus breaking the graph into pieces.
The example of Fig.\ \ref{convol} will help us explain the procedure.
It shows a particular embedding of $c_4$, how it is decomposed into two
pieces, and how its lattice constant can be computed accordingly. 
All we have to do is multiply the probability of $c_{2,c}$ (example 2)
with $m$ links at the bottom with the probability of $c_{m+2,b}$ (example 3), 
and then sum over $m$, resulting in
\beq
\langle c_4\rangle_x = \sum_{m=1}^{\infty}\frac{1}{2^{m-1}} (4- 2^{1-m}) = \frac{20}{3}.
\eeq
We can think of this operation as a product of an (infinite-dimensional) 
vector and matrix,
\beq\label{k1k2}
\begin{split}
K_1(m) &=\frac{1}{2^{m-1}},\\
K_2(m,n) &=\frac{1}{2^{m+n-1}}\matrice m+n \\ m \ematrice,
\end{split}
\eeq
followed by taking the trace in order to impose the open boundary condition,
that is,
\beq
\langle c_4\rangle_x = {\rm Tr} ( K_1\cdot K_2).
\eeq
Fig.\ \ref{convol} is a simple case with only two vertical lines in each strip. In general
there will be more lines, like in Fig.\ \ref{convol2}. 
However, there are only two possibilities for pairs of neighbouring vertical links:
they can be either disjoint or have exactly one vertex in common, 
so that the probability associated with them in a particular embedding in a 
particular triangulation will be given by either $K_2(m,n)$ or $K_1(n)$. 
We can then associate a probability to a given pattern of vertical lines in a strip with
fixed distance between them (in terms of the horizontal links in the in- and out-slice),
which is the product of the corresponding $K_1$- or 
$K_2$-probabilities, times a factor $\frac{1}{2}$ for each
shared vertical line between two neighbouring $K_i$-patterns in the same
strip (see Fig.\ \ref{convol2}).
In the end we have to glue the strips back together again, 
with conditions such as to get the desired graph, and
sum over the allowed values of horizontal lengths.
As an illustrative example, we obtain for Fig.\ \ref{convol2} the multiple sum
\beq \label{ex-stripdec}
\sum_{j_1=1}^{\infty}\sum_{m_1=1}^{\infty}\sum_{m_2=2}^{\infty}
\sum_{j_2=1}^{m_2-1}\sum_{n=1}^{\infty}
[K_2(l_1+j_1,m_1) \frac{1}{2} K_2(l_2,m_2)] [K_1(m_1) \frac{1}{2} K_2(m_2-j_2,n) \frac{1}{2} K_1(j_2)],
\eeq
where we have grouped the factors corresponding to the same strip in square brackets.
As should be clear from this example, the gluing and the range of the summations 
for the various graph embeddings
will have to be taken care of on a case-by-case basis.

Using these techniques we have repeated and independently 
confirmed the calculation of the first five orders given in \cite{aal1}
and then extended it to order 6. Table~\ref{tab-coeff} gives the lattice constants 
(times 2) per number of vertices (which we will
call the {\it susceptibility coefficients})
of the open chains, the other graphs, and of their grand total.
The effort involved in this extension is by no means minor.
There are 387 order-6 open chains that have to be counted individually, 
with the identifications introduced in this section,
and for many of them the computation requires a careful use of the strip decomposition and
a subsequent evaluation of the sums involved.

\begin{table}[hbtp]
\begin{center}
\begin{tabular}{|r||r|r|r|}
\hline
{\rule[-3mm]{0mm}{8mm} $n$} & \hspace{1cm} open & closed + disconnected & 
\hspace{.6cm} total ($a_n$)\\
\hline\hline
{\rule[-3mm]{0mm}{8mm} 1} & 6 & 0 & 6 \\
 \hline
{\rule[-3mm]{0mm}{8mm} 2} & 34 & 0 & 34 \\
\hline
{\rule[-3mm]{0mm}{8mm} 3} & 174 & 0 & 174 \\
 \hline
{\rule[-3mm]{0mm}{8mm} 4} & $859 \frac{1}{3}$ & $-12$ & $847 \frac{1}{3}$\\
 \hline
{\rule[-3mm]{0mm}{8mm} 5} & $4152 \frac{2}{3}$ & $-162 \frac{2}{3}$ & 3990\\
 \hline
{\rule[-3mm]{0mm}{8mm} 6} & $19800 \frac{19}{27}$ & $-1416 \frac{1}{9}$ & $18384 \frac{16}{27}$\\
  \hline
\end{tabular}
\caption{\footnotesize Susceptibility coefficients for the CDT lattice up to order 6.
The terminology `open', `closed' and `disconnected' refers to the topology 
of the graphs. }\label{tab-coeff}
\end{center}
\end{table}

\subsection{Counting graphs on the dual CDT lattice}

On the dual lattice (an example of which is depicted in Fig.\ \ref{cdt}, right-hand side) 
we can apply formula (\ref{recurr}), which for $q=3$ reduces to
\beq
\begin{split}
a_l-4\sigma a_{l-1} +4 a_{l-2} &= 2(l-2)d_{l-2}-2ld_l+ \\
&+8\left[m_l(1,1)+2m_{l-1}(1,1)+m_{l-2}(1,1)\right].
\end{split}
\eeq
Because of the low coordination number the only even graphs that can appear are 
connected
or disconnected polygons, and the only contributions to $m_l$ are from $\theta$-graphs 
and dumbbells
with or without disconnected polygonal components.
For the disconnected graphs we can use the reduction theorem. It should be noted that in the
overlap decomposition new topologies of graphs appear, for example, 
graphs with more than two odd vertices,
still rendering the higher-order computations non-trivial.

It is not difficult to derive the probability distribution relevant for the dual lattice 
from the one of the original lattice.
The probability of having two horizontal links at a vertex, one to the right and one to the
left, is again 1.
The probability for an incoming or outgoing vertical link at a vertex of a given 
chain of horizontal links is given by the probability of having a triangle or an upside-down 
triangle in the corresponding strip of the original triangulation. 
Since the two cases are mutually exclusive, the in- and out-probabilities are 
not independent and each takes the value $1/2$,  
which may also be thought of
as the weight associated with a triangle in the original lattice.
With these rules the lattice constant of an open chain of length 1 is $3/2$: 
1 for the horizontal
link (associated to its left vertex, say, in order to avoid overcounting), 
plus $1/2$ for the vertical link. Alternatively, 
this easy example can be computed by use of the Euler relation.
For higher-order graphs it is often convenient to go back to the original lattice 
and count -- with the appropriate
weights -- the configurations which can be associated to the dual graph under consideration.\\

\begin{figure}[ht]
\centering
\vspace*{12pt}
\includegraphics[width=9cm]{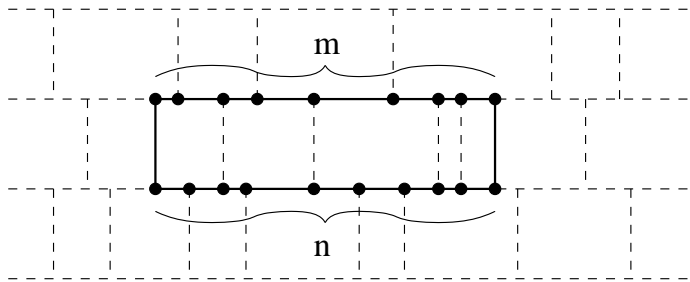}
\vspace*{13pt}
\caption{\footnotesize An embedding of the polygon considered in example 1.}
\label{dualpoly1}
\end{figure}

\noindent\textbf{Example 1.} Let us consider again a closed polygon with only two
vertical links, with $m$ links on the top and $n$ on the bottom, but this time 
on the dual CDT lattice (see Fig.\ \ref{dualpoly1}).
A careful evaluation of the probabilities involved yields 
\beq
\langle c_{m,n}\rangle^*=\frac{1}{2^{m+n+1}}\sum_{k=0}^{min(m,n)-1}\matrice m-1 \\ 
k \ematrice\matrice n-1 \\ k \ematrice 
\label{bino}
\eeq
for its lattice constant, where $k$ is the number of vertical links of the lattice which 
lie inside the polygon,
and $\langle\cdot\rangle^*$ denotes the average lattice constant per vertex of the
dual graph. 
The binomials in (\ref{bino}) count the number of possible ways to arrange the
$k$ internal links, and the remaining probability -- expressed as a power of
$1/2$ -- turns out not to depend on $k$.

As already mentioned, the in-and out-probabilities at a vertex of the dual lattice are 
not independent, which means that
we cannot apply the strip decomposition of the original lattice directly, but instead
have to proceed more carefully.\\

\begin{figure}[ht]
\centering
\vspace*{12pt}
\includegraphics[width=9cm]{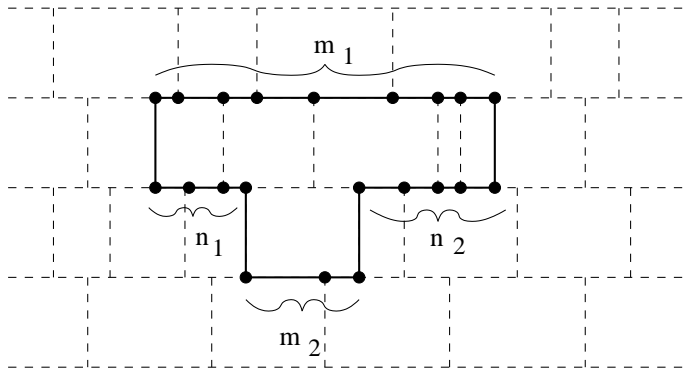}
\vspace*{13pt}
\caption{\footnotesize An embedding of the polygon considered in example 2.
}
\label{dualpoly2}
\end{figure}

\noindent\textbf{Example 2.} As an example of this, consider a polygon 
embedding that extends over
two strips (Fig.\ \ref{dualpoly2}), a type of configuration that appears from order 8 onward. Its lattice
constant is given by
\beq
\begin{split}
\langle c_{m_1,m_2,n_1,n_2}\rangle^* &=\sum_{k_1=0}^{m_1-1}\matrice m_1-1 \\ k_1 \ematrice
\sum_{k_2=0}^{m_2-1}\matrice m_2-1 \\ k_2 \ematrice \times \\
& \times \sum_{i=0}^{min(k_1,n_1+n_2-4)} \matrice k_1-i+k_2 \\ k \ematrice
\matrice n_1+n_2-2 \\ i \ematrice \frac{1}{2^{m_1+m_2+n_1+n_2+k_1-i+k_2+2}}.
\end{split}
\label{twostrip}
\eeq
In eq.\ (\ref{twostrip}), $m_1$ and $m_2$ count the links on the top and bottom horizontal 
lines, and $n_1$ and $n_2$ count the two sets of contiguous links on the central horizontal line.
The numbers of internal vertical links in the upper and lower strip are denoted by
$k_1$ and $k_2$, while $i$ counts how many links
out of the $k_1$ end on one of the two intermediate horizontal lines.
The logic behind the various counting factors appearing under the sums
is very similar to that of the previous example.

We have  found the method illustrated by examples 1 and 2 the most compact
one to deal with the counting of graph embeddings on the dual CDT lattice. It has
enabled us to push our ``counting by hand" to order 12, the results of which
are given in Table~\ref{tab-dual-coeff}. Because of the fixed, low coordination
number of the dual vertices, there are far fewer graphs at a given order than
there are on the original CDT lattice. 
\begin{table}[hbtp]
\begin{center}
\begin{tabular}{|r||r|r|r|r|r|r|r|r|r|r|r|r|}
\hline
{\rule[-3mm]{0mm}{8mm} $n$} & 1 & 2 & 3 & 4 & 5 & 6 & 7 & 8 & 9 & 10 & 11 & 12\\
\hline\hline
{\rule[-3mm]{0mm}{8mm} $a_n$} & 3 & 6 & 12 & 23 & 42+$\frac{3}{4}$ & 78+$\frac{1}{2}$ & 142+$\frac{3}{4}$
& 258 & 461+$\frac{13}{16}$ & 820+$\frac{1}{8}$ & 1446+$\frac{13}{32}$ & 2532+$\frac{11}{16}$\\
\hline
\end{tabular}
\caption{\footnotesize Susceptibility coefficients for the dual CDT lattice up to order 12.}
\label{tab-dual-coeff}
\end{center}
\end{table}

\section{Series analysis}

\subsection{Review of different methods}

Having computed the two high-temperature series to some order, we will now turn
to analyzing them, and try to extract information on the critical properties of the underlying
gravity-matter systems. The standard analytic methods are well-known from
the case of regular lattices and reviewed in 
\cite{guttmann}. We will recall them briefly here in order to be self-contained.

The first and most straightforward tool is that of the {\it ratio method}, which works as follows.
Assuming a simple behaviour of the susceptibility of the form
\beq
\chi (u)\sim A(u) \Bigl( 1-\frac{u}{u_c}\Bigr)^{-\gamma}+B(u)
\eeq
near the critical point $u_c$, with analytic functions $A$ and $B$, its series expansion $\chi (u)=
1+\sum_{n>0} a_n u^n$ should yield (to first order in $1/n$)
\beq\label{ratio}
r_n\equiv\frac{a_n}{a_{n-1}}=
\frac{1}{u_c} \Bigl( 1+\frac{\gamma - 1}{n}\Bigr).
\eeq
One can then generate sequences of estimates of the critical parameters from 
sequences of point pairs
$\{r_n,r_{n-1}\}$ \cite{guttmann}, namely,
\beq\label{seq1}
\gamma_n=\frac{n(2-n)r_n+(n-1)^2r_{n-1}}{n r_n-(n-1)r_{n-1}},
\eeq
\beq\label{seq2}
u_{c,n}=\frac{1}{n r_n-(n-1)r_{n-1}}.
\eeq
In general, these will converge very slowly. In addition, if the series has an
oscillatory behaviour the method can yield alternating over- and underestimates.
In this case a fit of a whole sequence $\{r_{min},...,r_{max}\}$, computed with
the help of (\ref{ratio}), may be better suited than the sequence of estimates
to find the straight line masked by the oscillations ($r_{min}$ is chosen
properly to exclude large deviations from (\ref{ratio}) at small $n$, and $r_{max}$ is the 
ratio computed for $n=n_{max}$).

Oscillations and other irregularities in the expansion are caused by a more 
complicated behaviour of the thermodynamic function, for example, 
the presence of other singularities in the complex plane close to the physical 
singularity, which in unfortunate cases may even lie closer to the origin than
the singularity of interest.
In case the behaviour near the physical singularity is like
\beq
\chi (u)\sim A(u) \Bigl( 1-\frac{u}{u_c}\Bigr)^{-\gamma},
\eeq
with $A(u)$ a {\it meromorphic} function with singularities close to $u_c$, 
a method that should give
better results than the ratio method is that of the so-called {\it Pad\'e approximants}.
The method consists in the approximation of a function, known through its series expansion 
to order $\cal N$, by a ratio of two polynomials of order $\cal L$ and $\cal M$, 
subject to the condition $\cal L+M\leq N$,
\beq\label{pade}
\frac{P_{\cal L}(x)}{Q_{\cal M}(x)}\equiv\frac{\sum_{k=0}^{\cal L} p_k x^k}
{1+\sum_{k=1}^{\cal M} q_k x^k}=F_{\cal N}(x)+O(x^{{\cal L+M}+1})
\eeq
with $F_{\cal N}(x)=\sum_{k=0}^{\cal N} a_k x^k$. By common usage, the notation 
$\cal [L/M]$ indicates 
the order of the polynomials used.

In our specific case one takes as the function $F(x)$ to be approximated 
the derivative of the logarithm of $\chi(u)$,
so that $u_c$ can be recognized as a pole and $\gamma$ as the associated residue
in
\beq
\frac{d}{du}\log \chi(u)=\frac{\gamma}{u_c-u}\left(1+O(u_c-u)\right).
\eeq
This is also referred to as the Dlog Pad\'e method.
Usually one only looks at the tridiagonal band $ [({\cal N}-1)/{\cal N}]$, $\cal [N/N]$ and 
$[({\cal N}+1)/{\cal N}]$ because of the invariance of the diagonal Pad\'e approximants 
under Euler transformations.

The Pad\'e approximants work well only when there is no additive term $B(u)$. 
In presence of such a term there is
a generalization -- known as {\it differential approximants} -- to account for 
functional behaviours of the form
\beq
\chi (u)\sim \prod_{i=1}^n A_i \Bigl( 1-\frac{u}{u_i}\Bigr)^{-\gamma_i} + B(u),
\eeq
with $B(u)$ and $A_1,...,A_n$ analytical functions and $u_1,...,u_n$ a set of singular points.
If we are considering the logarithmic derivative of a function $f(x)$
we can rewrite (\ref{pade}) as
\beq\label{dlog}
P_{\cal L}(x) f(x)-Q_{\cal M}(x) f'(x)=O(x^{{\cal L+M}+1}).
\eeq
The idea of the differential approximants method is to generalize this equation 
according to
\beq\label{diff-approx}
R_{{\cal M}_2}(x)F''_{\cal N}(x)+Q_{{\cal M}_1}(x) F'_{\cal N}(x)-P_{\cal L}(x) 
F_{\cal N}(x)=S_{\cal K}(x)+O(x^{{\cal K+L}+{\cal M}_1+{\cal M}_2+1}),
\eeq
where $R_{\mathcal{M}_2}(x)$ and $S_{\mathcal{K}}(x)$ are two other
polynomials
of order $\mathcal{M}_2$ and $\mathcal{K}$, and where
one can substitute the differential operator by $D=x\frac{d}{dx}$,
forcing the point at the origin to be a regular singular point.
In the following we will only consider the special case
$R_{\mathcal{M}_2}\equiv 0$,
giving rise to the so-called {\it inhomogeneous 1st-order
differential approximant}, denoted by
$[\mathcal{K}/\mathcal{L};\mathcal{M}_1]$.
With this method, the exponent $\gamma$ can be evaluated as
\beq
\gamma=\frac{P_{\mathcal{L}}(x_c)}{x_c Q'_{\mathcal{M}_1}(x_c)},
\label{gammada}
\eeq
where $x_c$ is a simple root\footnote{The case of multiple roots can also
be considered, but in our case never occurs.}
of $Q_{\mathcal{M}_1}(x)$, which gives an estimate
of the critical point.

\subsection{The CDT lattice series}

\begin{figure}
\centering
\includegraphics[width=11cm]{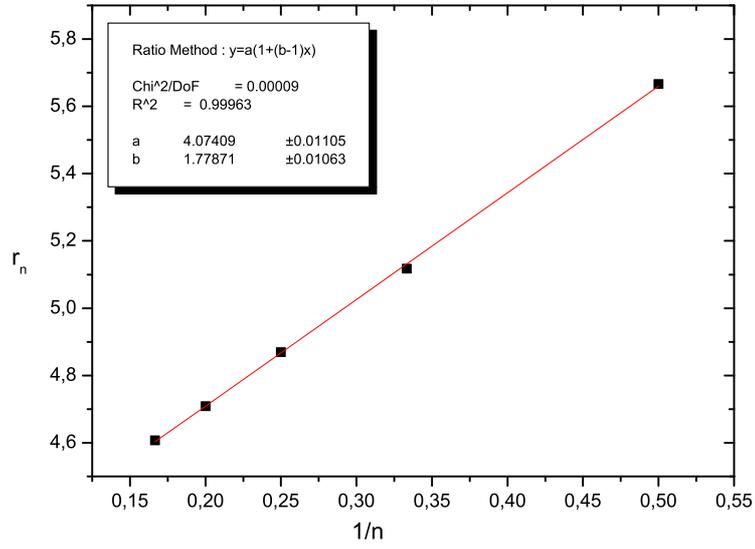}
\caption{\footnotesize Plot of the ratios (\ref{ratio}) for the case of the CDT lattice
(data from Table~\ref{tab-fits}).}
\label{ratioplot1}
\end{figure}

The high-temperature series expansion for the Ising susceptibility on the CDT model is
given by
\beq \label{h-series}
\chi_{\rm CDT}(u)=1+6u+34u^2+174u^3+(847+\frac{1}{3})u^4+3990u^5+
(18384+\frac{16}{27})u^6+O(u^7).
\eeq
Applying formulas (\ref{seq1}) and (\ref{seq2}) we get the sequence of estimates for 
$\gamma$ and $u_c$ reported in Table~\ref{tab-2ptseq}.
\begin{table}[hbtp]
\begin{center}
\begin{tabular}{|c||c|c|}\hline
{\rule[-3mm]{0mm}{8mm} $n$ }
& \hspace{1cm} $u_c \hspace{1cm}$ & \hspace*{25pt} $\gamma$ \hspace*{25pt}\\ \hline\hline
{\rule[-3mm]{0mm}{8mm} 3} & 0.2489 &
1.819\\ \hline
{\rule[-3mm]{0mm}{8mm} 4} & 0.2424 &
1.721\\ \hline
{\rule[-3mm]{0mm}{8mm} 5} & 0.2459 &
1.791\\  \hline
{\rule[-3mm]{0mm}{8mm} 6} & 0.2438 &
1.740\\  \hline
\end{tabular}
\caption{\footnotesize Two-point linear extrapolations of the critical point and the critical exponent
of the series (\ref{h-series}) obtained using  (\ref{seq1}) and (\ref{seq2}).}\label{tab-2ptseq}
\end{center}
\end{table}
The results of linearly fitting the whole sequence of ratios $\{r_2,...,r_{max}\}$ instead 
are listed in Table~\ref{tab-fits}, 
and a plot of the ratios $\{r_2,...,r_6\}$ versus $1/n$ together with their fit is shown in 
Fig.\ \ref{ratioplot1}. Despite the shortness of the sequence, the linear fit is of very
good quality.
\begin{table}[hbtp]
\begin{center}
\begin{tabular}{|c||c|c|}\hline
{\rule[-3mm]{0mm}{8mm} $n_{\rm max}$ }
& \hspace{1cm} $u_c$  \hspace{1cm} & \hspace*{25pt} $\gamma$ \hspace*{25pt}\\
\hline\hline
{\rule[-3mm]{0mm}{8mm} 3} & \hspace{.2cm} 0.2488
\hspace{.2cm} &
1.820\\ \hline
{\rule[-3mm]{0mm}{8mm} 4} & 0.2462 &
1.789\\ \hline
{\rule[-3mm]{0mm}{8mm} 5} & 0.2458 &
1.783\\  \hline
{\rule[-3mm]{0mm}{8mm} 6} & 0.2454 &
1.779\\  \hline
\end{tabular}
\caption{\footnotesize Linear fits of the sequences $\{r_2,...,r_{max}\}$ for the series 
(\ref{h-series}) with assumed functional form (\ref{ratio}).}\label{tab-fits}
\end{center}
\end{table}
It is worth noting that if we remove the point $r_2$, which of course is subject
to the largest deviations from a pure $1/n$-behaviour due to higher-order terms, 
and fit the truncated sequence $\{r_3,...,r_6\}$, we obtain $\gamma=1.749$ and 
$u_c=0.2441$ (see Table~\ref{tab-fits2}).

\begin{figure}[ht]
\centering
\includegraphics[width=11cm]{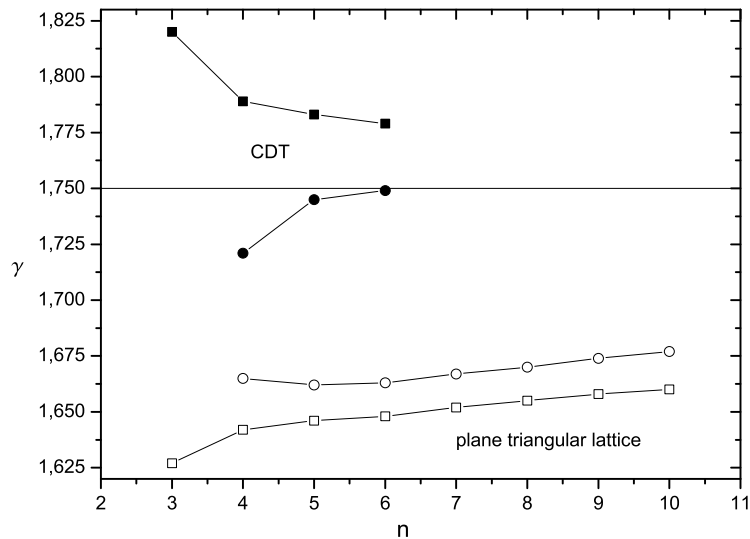}
\caption{\footnotesize A comparative plot of estimates of the critical exponent 
$\gamma$ for the Ising model on the CDT versus that on a plane triangular lattice
(data taken from \cite{sykes-h}), obtained via the ratio method. 
The filled shapes refer to the CDT model and the open
ones to the plane triangular lattice.
The squares indicate fits of $\{r_{n_{min}},...,r_n\}$ sequences, and the circles
those of sequences $\{r_{n_{min}+1},...,r_n\}$ with the first ratio eliminated.}
\label{gamma}
\end{figure}

\begin{table}[hbtp]
\begin{center}
\begin{tabular}{|c||c|c|}\hline
{\rule[-3mm]{0mm}{8mm} $n_{\rm max}$ }
& \hspace{1cm} $u_c$  \hspace{1cm} & \hspace*{25pt} $\gamma$ \hspace*{25pt}\\
\hline\hline
{\rule[-3mm]{0mm}{8mm} 4} & 0.2424 &
1.721\\ \hline
{\rule[-3mm]{0mm}{8mm} 5} & 0.2439 &
1.745\\  \hline
{\rule[-3mm]{0mm}{8mm} 6} & 0.2441 &
1.749\\  \hline
\end{tabular}
\caption{\footnotesize Linear fits of the sequences $\{r_3,...,r_{max}\}$ for the 
series (\ref{h-series}) with assumed
functional form (\ref{ratio}).}\label{tab-fits2}
\end{center}
\end{table}
It is surprising how fast the ratio sequence seems to converge to the result 
$\gamma=1.75$ of the exact solution for the flat regular case,
as compared to the analogous one obtained from the series expansion on the plane 
triangular lattice (see \cite{sykes-h} for the expansion coefficients to order $u^{16}$), even
though we cannot claim the result to be conclusive because of the limited number of 
terms at our disposal (see Fig.\ \ref{gamma}).

What is the picture when we use one of the alternative methods to evaluate the
series expansions? Results from the Dlog Pad\'e and differential approximants 
methods are somewhat inconclusive, most likely because of the small number of 
estimates we can perform with our 6th-order series (see Table~\ref{tab-pade}).
To get a feel for what may be expected at this order, we report in Table~\ref{tab-pt} 
the corresponding Dlog Pad\'e approximants $[({\cal N}+j)/\cal N]$ to the series for the
plane triangular lattice (see again \cite{sykes-h}). The estimates for the critical
exponent $\gamma$ in the latter case are clearly closer to the known exact value,
whereas for the CDT model, with only two values for the
diagonal $\cal [N/N]$, it is quite impossible to extrapolate the behaviour of $\gamma$.  
(Note that we are reporting the estimates for the critical point $u_c$ only for
completeness; they are or course not expected to coincide for the two models.)
We have also computed the inhomogeneous first-order differential
approximants $[{\cal L}/({\cal N}+j);\cal N]$ for the CDT series, which was defined
just before formula (\ref{gammada}). Values for the critical
exponents are not completely off,  but there are simply too few of them to make
any statement about their convergence behaviour (see Table~\ref{tab-diff}).

\begin{table}[hbtp]
\begin{center}
\begin{tabular}{|c||c|c|c|c|c|c|}\hline   &
\multicolumn{2}{|c|}{{\rule[-3mm]{0mm}{8mm}
$[({\cal N}-1)/{\cal N}]$}} & \multicolumn{2}{|c|}{{\rule[-3mm]{0mm}{8mm}
$[{\cal N}/{\cal N}]$}} & \multicolumn{2}{|c|}{{\rule[-3mm]{0mm}{8mm}
$[({\cal N}+1)/{\cal N}]$}} \\
\hline
{\rule[-3mm]{0mm}{8mm} ${\cal N}$} & $u_c$ & $\gamma$ & $u_c$ & $\gamma$ & $u_c$ & $\gamma$ \\
\hline\hline
{\rule[-3mm]{0mm}{8mm} 1} & 0.1875 & 1.125 & 0.2540 & 2.064 & 0.2513 & 2.000 \\
\hline
{\rule[-3mm]{0mm}{8mm} 2} & 0.2514 & 2.003 & 0.2483 & 1.8810 & 0.2517 & 2.008 \\
\hline
{\rule[-3mm]{0mm}{8mm} 3} & 0.2519 & 2.012 & - & - & - & - \\
\hline
\end{tabular}
\caption{\footnotesize Dlog Pad\'e approximants method applied to the CDT 
series (\ref{h-series}).}\label{tab-pade}
\end{center}
\end{table}
\begin{table}[hbtp]
\begin{center}
\begin{tabular}{|c||c|c|c|c|c|c|}\hline   &
\multicolumn{2}{|c|}{{\rule[-3mm]{0mm}{8mm}
$[({\cal N}-1)/{\cal N}]$}} & \multicolumn{2}{|c|}{{\rule[-3mm]{0mm}{8mm}
$[{\cal N}/{\cal N}]$}} & \multicolumn{2}{|c|}{{\rule[-3mm]{0mm}{8mm}
$[({\cal N}+1)/{\cal N}]$}} \\
\hline
{\rule[-3mm]{0mm}{8mm} ${\cal N}$} & $u_c$ & $\gamma$ & $u_c$ & $\gamma$ & $u_c$ & $\gamma$ \\
\hline\hline
{\rule[-3mm]{0mm}{8mm} 1} & 0.2500 & 1.500 & 0.2666 & 1.706 & 0.2678 & 1.730 \\
\hline
{\rule[-3mm]{0mm}{8mm} 2} & 0.2679 & 1.732 & 0.2670 & 1.712 & 0.2661 & 1.688 \\
\hline
{\rule[-3mm]{0mm}{8mm} 3} & 0.2662 & 1.692 & 0.2667 & 1.705 & 0.2672 & 1.722 \\
\hline
\end{tabular}
\caption{\footnotesize Dlog Pad\'e approximants method applied to the plane triangular 
series in \cite{sykes-h}.}\label{tab-pt}
\end{center}
\end{table}

\begin{table}[hbtp]
\begin{center}
\begin{tabular}{|c||c||c|c|}\hline   & &
\multicolumn{2}{|l|}{{\rule[-3mm]{0mm}{8mm}
${\cal N}=$}}  \\
{\rule[-3mm]{0mm}{8mm} ${\cal L}$} & $j$ & 1 & 2 \\
\hline\hline
1 & -1 & $u_c=0.2143$ & \hspace*{3mm} 0.2454 \hspace*{3mm} \\
  &  & $\gamma=1.767$ & 1.815 \\
  & 0 & 0.2373 & 0.2461 \\
  &  & 1.568 & 1.832 \\
  & 1 & 0.2496 & - \\
  &  & 1.944 & - \\
\hline
2 & -1 & 0.2252 & 0.2463\\
  &  & 1.265 & 1.840 \\
  & 0 & 0.2529 & - \\
  &  & 2.087 & - \\
  & 1 & 0.2431 & - \\
  &  & 1.685 & - \\
\hline
3 & -1 & 0.2334 & - \\
  &  & 1.361 & - \\
  & 0 & 0.2402 & - \\
  &  & 1.577 & - \\
  & 1 & - & - \\
  &  & - & - \\
\hline
4 & -1 & 0.2357 & - \\
  &  & 1.396 & - \\
  & 0 & - & - \\
  &  & - & - \\
  & 1 & - & - \\
  &  & - & - \\
\hline
\end{tabular}
\caption{\footnotesize Inhomogeneous 1st-order differential approximants 
$[{\cal L}/({\cal N}+j);{\cal N}]$ method for the  series (\ref{h-series}).}
\label{tab-diff}
\end{center}
\end{table}

\clearpage

\subsection{The dual CDT lattice series}

The high-temperature series expansion for the Ising susceptibility on the dual CDT model is
given by
\beq \label{h-dual-series}
\begin{split}
\chi_{\rm CDTd}(u) &=1+3u+6u^2+12u^3+23u^4+(42+\frac{3}{4})u^5+(78+\frac{1}{2})u^6+(142+\frac{3}{4})u^7\\
   +&258u^8+(461+\frac{13}{16})u^9+(820+\frac{1}{8})u^{10}+(1446+\frac{13}{32})u^{11}+(2532+\frac{11}{16})u^{12}+O(u^{13})
\end{split}
\eeq
Using expressions (\ref{seq1}) and (\ref{seq2}) we get the sequence of ratio-method 
estimates for $\gamma$ and $u_c$ reported in Table~\ref{tab-dual-2ptseq}.
\begin{table}[hbtp]
\begin{center}
\begin{tabular}{|c||c|c|}\hline
{\rule[-3mm]{0mm}{8mm} $n_{\rm max}$ }
& \hspace{1cm} $u_c \hspace{1cm}$ & \hspace*{25pt} $\gamma$ \hspace*{25pt}  \\ \hline\hline
{\rule[-3mm]{0mm}{8mm} 3} & 0.5 &
1.\\ \hline
{\rule[-3mm]{0mm}{8mm} 4} & 0.6 &
1.6\\ \hline
{\rule[-3mm]{0mm}{8mm} 5} & 0.6147 &
1.713\\  \hline
{\rule[-3mm]{0mm}{8mm} 6} & 0.5800 &
1.390\\  \hline
{\rule[-3mm]{0mm}{8mm} 7} & 0.5842 &
1.436\\ \hline
{\rule[-3mm]{0mm}{8mm} 8} & 0.5782 &
1.360\\ \hline
{\rule[-3mm]{0mm}{8mm} 9} & 0.6057 &
1.758\\  \hline
{\rule[-3mm]{0mm}{8mm} 10} & 0.6064 &
1.769\\  \hline
{\rule[-3mm]{0mm}{8mm} 11} & 0.6093 &
1.820\\  \hline
{\rule[-3mm]{0mm}{8mm} 12} & 0.6202 &
2.033\\  \hline
\end{tabular}
\caption{\footnotesize Two-point linear extrapolations of the critical point and the critical exponent
of the dual series (\ref{h-dual-series}) obtained using (\ref{seq1}) and (\ref{seq2}).}
\label{tab-dual-2ptseq}
\end{center}
\end{table}

\begin{figure}
\centering
\includegraphics[width=11cm]{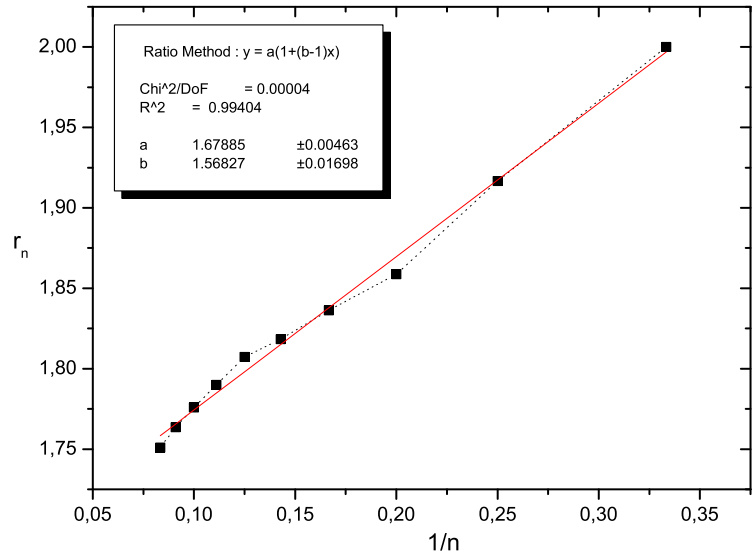}
\caption{\footnotesize Plot of the ratios (\ref{ratio}) for the case of the dual CDT lattice.}
\label{ratioplot2}
\end{figure}

\begin{figure}
\centering
\includegraphics[width=11cm]{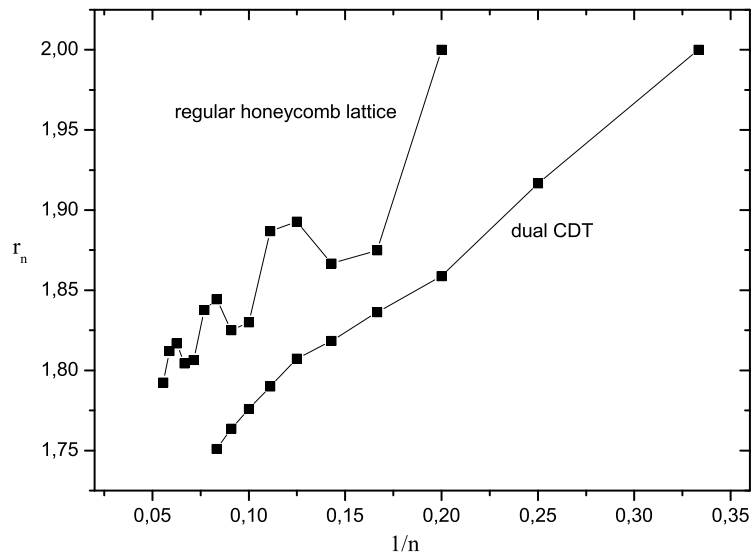}
\caption{\footnotesize A comparative plot of the ratios of the regular honeycomb 
lattice (data taken from \cite{sykes-h})
versus the dual CDT model.}
\label{HCvsDual}
\end{figure}

\noindent The relative ratio plot is shown in Fig.\ \ref{ratioplot2}, and shows gentle
oscillations around the best linear fit of the sequence 
$\{ r_3,...,r_{12}\}$, leading to the estimates $\gamma=1.568$ and $u_c=0.5956$.
However, these numbers do not tell it all. The strong qualitative differences with
the planar honeycomb lattice, the appropriate non-random
version of the dual CDT lattice, are illustrated in Fig.\ \ref{HCvsDual}. It shows a
comparative plot of the ratios extracted for the two models (the expansion coefficients
for the honeycomb lattice, known to order $u^{32}$, can be found in \cite{sykes-h}).
Obviously, the influence of interfering unphysical singularities present on the
regular honeycomb lattice is much reduced in the fluctuating dual CDT ensemble.


On the basis of the rather well-behaved results for the Ising susceptibility
on dynamically triangulated spacetimes we conjectured in \cite{ising1} 
that ``coupling to quantum gravity" may be an optimal method to learn
about the critical behaviour of a matter or spin system, in the sense that
the randomness of the underlying geometry eliminates spurious unphysical
singularities (due to the presence of discrete symmetries in the case of
regular lattices), but is not strong enough to change the universality
class of the matter system. This is a dynamical version of similar
conjectures made in the 1980s, which advocated the use of {\it fixed}
random lattices to improve the convergence behaviour of lattice gauge theory,
say (see, for example, \cite{random}), but ultimately did not succeed.

If it is indeed the case that the singularity structure of the thermodynamic
quantities in the complex plane is simplified, this could explain that the
simple ratio method does indeed give the best result, at least at the relatively
low orders we have been considering. In the case of the Ising model coupled
to Euclidean dynamical triangulations, an analogous result has already been 
obtained with regard to the locus of the zeros of the partition function 
in the complex-temperature plane, where unphysical singularities
of the flat regular lattices have been shown to be largely absent \cite{fat}.
However, as already mentioned, the
underlying geometries are too disordered to serve our present purpose, because
their critical matter behaviour is changed compared to the flat case.    \\

\begin{table}[hbtp]
\begin{center}
\begin{tabular}{|c||c|c|c|c|c|c|}\hline   &
\multicolumn{2}{|c|}{{\rule[-3mm]{0mm}{8mm}
$[({\cal N}-1)/{\cal N}]$}} & \multicolumn{2}{|c|}{{\rule[-3mm]{0mm}{8mm}
$[{\cal N}/{\cal N}]$}} & \multicolumn{2}{|c|}{{\rule[-3mm]{0mm}{8mm}
$[({\cal N}+1)/{\cal N}]$}} \\
\hline
{\rule[-3mm]{0mm}{8mm} ${\cal N}$} & $u_c$ & $\gamma$ & $u_c$ & $\gamma$ & $u_c$ & $\gamma$ \\
\hline\hline
{\rule[-3mm]{0mm}{8mm} 3} & 0.5952 & 1.554 & 0.5574 & 0.9738 & 0.5686 & 1.170 \\
\hline
{\rule[-3mm]{0mm}{8mm} 4} & 0.5691 & 1.180 & 0.5564 & 0.9567 & 0.5861 & 1.446 \\
\hline
{\rule[-3mm]{0mm}{8mm} 5} & 0.5890 & 1.492 & 0.5935 & 1.560 & 0.6359 & 2.996 \\
\hline
{\rule[-3mm]{0mm}{8mm} 6} & 0.5761 & 1.402 & - & - & - & - \\
\hline
\end{tabular}
\caption{\footnotesize Dlog Pad\'e approximants method for the dual CDT series (\ref{h-dual-series}).}\label{tab-dual-pade}
\end{center}
\end{table}
\begin{table}[hbtp]
\begin{center}
\begin{tabular}{|c||c|c|c|c|c|c|}\hline   &
\multicolumn{2}{|c|}{{\rule[-3mm]{0mm}{8mm}
$[({\cal N}-1)/{\cal N}]$}} & \multicolumn{2}{|c|}{{\rule[-3mm]{0mm}{8mm}
$[{\cal N}/{\cal N}]$}} & \multicolumn{2}{|c|}{{\rule[-3mm]{0mm}{8mm}
$[({\cal N}+1)/{\cal N}]$}} \\
\hline
{\rule[-3mm]{0mm}{8mm} ${\cal N}$} & $u_c$ & $\gamma$ & $u_c$ & $\gamma$ & $u_c$ & $\gamma$ \\
\hline\hline
{\rule[-3mm]{0mm}{8mm} 3} & - & - & - & - & 0.6063 & 2.153 \\
\hline
{\rule[-3mm]{0mm}{8mm} 4} & 0.5589 & 1.416 & 0.5680 & 1.531 & 0.5676 & 1.525 \\
\hline
{\rule[-3mm]{0mm}{8mm} 5} & 0.5676 & 1.525 & 0.5679 & 1.530 & 0.5589 & 1.459 \\
\hline
{\rule[-3mm]{0mm}{8mm} 6} & 0.5645 & 1.490 & 0.5730 & 1.618 & 0.5709 & 1.571 \\
\hline
\end{tabular}
\caption{\footnotesize Dlog Pad\'e approximants method for the honeycomb lattice series 
in \cite{sykes-h}.}
\label{tab-hc}
\end{center}
\end{table}

\begin{table}[hbtp]
\begin{center}
\begin{tabular}{|c||c||c|c|c|}\hline   & &
\multicolumn{3}{|l|}{{\rule[-3mm]{0mm}{8mm}
${\cal N}=$}}  \\
{\rule[-3mm]{0mm}{8mm} $\cal L$} & $j$ & 2 & 3 & 4 \\
\hline\hline
1 & -1 & $u_c=0.5000$ & \hspace*{3mm} 0.6246 \hspace*{3mm}& \hspace*{3mm} 0.5873 \hspace*{3mm}\\
  &  & $\gamma=1.000$ & 2.1435 & 1.503\\
  & 0 & 0.4974 & 0.5722 & 0.4797\\
  &  & 1.018 & 1.205 & 2.688\\
  & 1 & 0.5782 & 0.5759 & 0.6024\\
  &  & 1.298 & 1.234 & 1.828\\
\hline
2 & -1 & 0.4967 & 0.5767 & 0.6094\\
  &  & 1.023 & 1.312 & 1.932\\
  & 0 & 0.5735 & 0.5802 & 0.5768\\
  &  & 1.265 & 1.395 & 1.029\\
  & 1 & 0.5753 & 0.5620 & 0.4902\\
  &  & 1.276 & 0.9503 & 2.201\\
\hline
3 & -1 & 0.5722 & 0.5810 & 0.5966\\
  &  & 1.216 & 1.414 & 1.622\\
  & 0 & 0.5715 & 5692 & 0.5010\\
  &  & 1.195 & 1.208 & 2.031\\
  & 1 & 0.5820 & 0.5606 & - \\
  &  & 1.436 & 0.9257 & - \\
\hline
4 & -1 & 0.5735 & 0.5502 & 0.4711\\
  &  & 1.237 & 1.062 & 2.966\\
  & 0 & 0.5867 & 0.5100 & - \\
  &  & 1.571 & 1.880 & - \\
  & 1 & 0.5376 & 0.6467 & - \\
  &  & 1.202 & 3.062 & - \\
\hline
5 & -1 & 0.5882 & complex & - \\
  &  & 1.572 & complex & - \\
  & 0 & complex & - & - \\
  &  & complex & - & - \\
  & 1 & complex & - & - \\
  &  & complex & - & - \\
\hline
\end{tabular}
\caption{\footnotesize Inhomogeneous 1st order differential approximants 
$[{\cal L}/{\cal N}+j;{\cal N}]$ method for dual CDT
series (\ref{h-dual-series}).}
\label{tab-dual-diff}
\end{center}
\end{table}
\noindent Returning to the evaluation of the
dual CDT series, we have used both the Dlog Pad\'e and the differential 
approximants methods, in addition to the ratio method already described.  
Even with the larger number of terms compared to the original CDT case
the results are reluctant to show any clear sign of convergence. Our results
for the Dlog Pad\'e approximant are summarized in Table~\ref{tab-dual-pade}.
The corresponding data for the regular honeycomb lattice \cite{sykes-h},
which we are including for comparison in Table~\ref{tab-hc}, seem to show more
consistency at the same order of approximation. 
The result from using the
inhomogeneous first-order differential approximants for the dual CDT lattice
(see Table~\ref{tab-dual-diff}) is similarly inconclusive. It would clearly be
desirable to understand in more analytic terms the influence of the 
fluctuating geometry of the underlying quantum gravitating lattice on the
singularity structure of spin models like the one we have been
considering, and thus determine which of the approximation methods is
best suited for extracting their critical behaviour.

\section{Comments on the low-temperature expansion}

Having dealt with the high-temperature expansion, it is natural to ask whether similar
expansion techniques can be applied in a low-temperature expansion of our
coupled models of matter and gravity, in the same way this is possible for
spin systems on regular lattices. The first thing to notice is that at the point of
zero temperature around which one expands, the spins are all frozen
to point into the same direction, and effectively play no role. The 
quantum-gravitational ensemble is therefore 
again characterized by the probability distribution (\ref{vertprob}) for the coordination number
of pure gravity, as was also the case in the limit of infinite temperature.

In the low-temperature expansion on the regular lattice 
one starts from a ground state with all spins
aligned, and then includes perturbations with $s$ overturned spins,
obtaining \cite{sykes-l}
\beq\label{low-exp}
  \ln Z_N(K,H)=l K + v H + \sum_{s,r} [ s;r;G ] z^p \mu^s,
\eeq
in terms of the expansion parameters $z=e^{-2K}$ and $\mu=e^{-2H}$, where
$K$, $H$, $l$ and $v$ were all defined at the beginning of Sec.\ 3 above. 
The part linear in $N$ of the (strong) lattice constant (see \cite{domb1} for a
definition) for a graph
with $s$ vertices and $r$ links embedded in $G$ is denoted by 
$[s;r;G]$, and $p$ is the number of lattice links 
which are incident on any of the vertices of the graph, but do not themselves belong to 
the embedded graph.
On a lattice of fixed coordination number $q$ it is easy to prove that $p=qs-2r$. 
Graphs with given values of $s$ and $r$ therefore contribute only at a specific order.

The low-temperature expansion on CDT lattices via graph counting
represents a slight complication. It is easy to see that already by overturning just
a single spin the number of links whose interaction changes sign as a result depends on the 
spin's position, so that
each site will contribute to a different order in the perturbative 
expansion.\footnote{By contrast, for the low-temperature expansion on {\it dual} CDT 
lattices, where the coordination number is fixed to 3, the expansion parallels
that on flat lattices, with the (strong) lattice constants replaced by their 
ensemble averages.} 
In the case at hand, this difficulty
can be overcome by counting  the number of elementary
polygons (i.e. faces) with given boundary length $p$ {\it on the dual lattice}. 
In the case of several overturned spins we can proceed similarly by associating
certain patterns on the dual lattice with a given order of $z$. 
More precisely, to obtain the coefficient at order $z^p \mu^s$ we need to count 
polygons on the dual lattice which have
total boundary length $p$ and are constructed out of $s$ faces. 
Denoting the number of such patterns
by $\mathcal{P}(p,s)$, we can write the free energy per unit volume as
\beq\label{low-exp2}
  \frac{1}{N}\ln Z_N(K,H)=\frac{3}{2} K + \frac{1}{2} H + \sum_{s,p} \mathcal{P}(p,s) z^p \mu^s
\eeq
and the susceptibility at vanishing magnetic field as
\beq \label{low-chi}
  \chi(z)=\sum_{s,p} 4 s^2 \mathcal{P}(p,s) z^p.
\eeq
Although the procedure is slightly more involved, and cannot be seen as a 
computation of (averaged) strong lattice constants, the expansion is still doable 
using the counting techniques introduced earlier in this paper.
Table~\ref{tab-low-coeff} gives the results for $\mathcal{P}(p,s)$ up to order 10.
\begin{table}[hbtp]
\begin{center}
\begin{tabular}{|c||c|c|c|c|}\hline
{\rule[-3mm]{0mm}{8mm}  } &  $s=1$ & $s=2$ & $s=3$ & $s=4$ \\ \hline\hline
{\rule[-3mm]{0mm}{8mm} $p=4$} & $\frac{1}{8}$ & - & - & - \\ \hline
{\rule[-3mm]{0mm}{8mm} $p=5$} & $\frac{1}{8}$ & - & - & - \\ \hline
{\rule[-3mm]{0mm}{8mm} $p=6$} & $\frac{3}{32}$ & $\frac{1}{32}$ & - & - \\ \hline
{\rule[-3mm]{0mm}{8mm} $p=7$} & $\frac{1}{16}$ & $\frac{1}{16}$ & - & - \\ \hline
{\rule[-3mm]{0mm}{8mm} $p=8$} & $\frac{5}{128}$ & $\frac{3}{64}$ & $\frac{1}{128}$ & - \\ \hline
{\rule[-3mm]{0mm}{8mm} $p=9$} & $\frac{3}{128}$ & $\frac{9}{64}$ & $\frac{7}{128}$ & - \\ \hline
{\rule[-3mm]{0mm}{8mm} $p=10$} & $\frac{7}{512}$ & $\frac{19}{512}$ & $\frac{29}{512}$ & $\frac{9}{512}$ \\ \hline
\end{tabular}
\caption{\footnotesize Free energy coefficients $\mathcal{P}(p,s)$ as defined in (\ref{low-exp2})
for the CDT lattice.}\label{tab-low-coeff}
\end{center}
\end{table}
The susceptibility low-temperature series (\ref{low-chi}) turns out to be very irregular 
and none of the methods of analysis considered seems to give
a reasonable indication of the critical exponent, at least not from the relatively few terms 
we have computed. Although this is somewhat disappointing, in view of the fact that 
low-temperature expansions for regular lattices are notoriously ill-behaved, 
it does not come as a total surprise.

What helps in the evaluation of the low-temperature series for
the Ising model on {\it flat} triangular, square or honeycomb lattices is the fact that
their critical temperatures are known exactly from duality arguments 
and thus can be used as an input
in so-called biased approximants to considerably improve the estimates of
critical exponents. Namely, for
$H=0$ one can apply the standard duality transformation \cite{baxter} which maps
the low-temperature expansion (\ref{low-exp}) of the triangular lattice model to 
the high-temperature
expansion (\ref{high-exp}) of its dual and vice versa.
This can be combined with the star-triangle transformation to obtain
the critical point. The analogous deriviation for the simple square lattice
is even easier, because it is self-dual.
Unfortunately, the star-triangle transformation is not applicable in the CDT case,
which prevents us from making a similar argument in the coupled Ising-gravity
model. In order to pursue the analysis of the low-temperature series further,
we would therefore have to rely on the evaluation of higher-order terms in
the expansion.

\section{Conclusions and outlook}

In this article we have given a concrete example of a method which can
be used to extract physical properties of a strongly coupled model of
gravity and matter.
We showed how the method -- estimating critical matter exponents from
a series expansion of suitable thermodynamic functions of the system --
can be adapted successfully from the case of a fixed, flat lattice to that
of a fluctuating ensemble of geometries, as is relevant in studies of
non-perturbative quantum gravity. For the ensemble of two-dimensional
causal dynamical triangulations, we have formulated an explicit algorithm for
counting embedded graphs which allows us to do the counting
recursively for increasing order. The method can in principle be
applied to other matter and spin systems which admit a similar
diagrammatic expansion in terms of weak or strong lattice constants
around infinite or zero
temperature, and where one has sufficient information about the
probability distribution of local geometric lattice properties like the
vertex coordination number. Even in cases where these lattice
properties are not available explicitly, the lattice constants could still be
extracted from simulating the pure gravity ensemble.

As a potential spin-off, we noticed that the explicit expansion for the
susceptibility in vanishing external field of the Ising model on CDT (up
to order 6) or
dual CDT lattices (up to order 12) indicates a more regular behaviour of
this function
in the complex-temperature plane than that of the Ising model on the
corresponding fixed triangular or honeycomb lattice. This may also
explain why the simple ratio method, applied to CDT lattice results from
only six orders, gives an excellent approximation to the Onsager
susceptibility exponent $\gamma=1.75$. As we saw in some detail, other
approximation methods, namely, Dlog Pad\'e and differential approximants,
do not produce results of a similar quality. We believe that for the case of
the dual CDT lattice, the series computed is simply too short to yield
reliable estimates with any of the approximation methods
(even at order 12, the number of terms contributing
is much smaller than the number of terms contributing at order 6 on
the original CDT lattices). At any rate, if one aims to make an argument of
improved convergence of matter behaviour on {\it fluctuating} lattices
(see also \cite{ising1}),
it is plausible that this will be achieved optimally with the triangulated
CDT geometries
whose vertex coordination number can vary dynamically from 4 all
the way to infinity, rather than with the dual tesselations which have
a fixed, low coordination number of 3. To investigate the convergence
issue in more detail will
require going to higher order than 6 on CDT lattices, which is not really
feasible `by hand', as we have been doing so far, but will require
the setting up of a computer algorithm to perform (at least part of)
the graph counting.

Our work should be seen as contributing to the study of non-perturbative
systems of quantum gravity coupled to matter, which is only just beginning.
One challenging task will be
to establish {\it computable} criteria characterizing and quantifying
the influence of geometry on matter and vice versa, in the physically
relevant case of four spacetime dimensions, something about which
we currently know close to nothing. Obvious quantities of interest 
are critical exponents pertaining to geometry (like, for instance, the
Hausdorff and spectral dimensions of spacetime already measured for
the ground state of four-dimensional CDT \cite{ajl-rec}), and critical
matter exponents, like that of the susceptibility investigated in the present
work. One would like to have a classification of possible universality classes 
of gravity-matter models as a function of the characteristics of the
ensemble of quantum-fluctuating geometries. A necessary condition
for viable quantum gravity models is that at sufficiently large distances
and for sufficiently weak matter fields the correct classical limits for
these critical parameters must be recovered. 

Two-dimensional toy models like the one we have been considering
can serve as a blueprint for what phenomena one might expect to
find. In two dimensions, there have been several investigations of the effect 
of random geometric disorder on the critical properties of matter systems,
and attempts to formulate general criteria for when a particular type of
disorder is relevant, i.e. will lead to a matter behaviour different from that
on fixed, regular lattices. Good examples are the Harris \cite{harris} and the 
Harris-Luck \cite{harrisluck} criteria,
which tie the relevance of random disorder to the
value of the specific-heat exponent $\alpha$ and to correlations among 
the disorder degrees of freedom. However, not all models fit the predictions,
and open problems remain (see \cite{luck} and references therein).
Causal dynamical triangulations coupled to matter add another class of 
models inspired by quantum gravity,
whose randomness lies in between that of Poissonian Voronoi-Delaunay
triangulations and the highly fractal Euclidean dynamical triangulations,
and over which 
one has some analytic control. It will be interesting to see to what degree
the robustness of the matter behaviour with respect to the geometric 
fluctuations observed so far \cite{aal1,aal2,ising1} will persist for
different types of spin and matter systems and in higher dimensions.

\vspace{.7cm}

\noindent {\bf Acknowledgements.} We thank F.T.\ Lim for discussions during
the early stages of the project. Both authors are partially supported 
through the European Network on Random Geometry
ENRAGE, contract MRTN-CT-2004-005616. R.L. acknowledges 
support by the Netherlands Organisation for Scientific Research 
(NWO) under their VICI program.

\bibliographystyle{unsrt}

\end{document}